\documentclass[aps,prl,reprint,superscriptaddress]{revtex4-1}

\usepackage{graphicx}
\usepackage{subfigure}
\usepackage{float}
\usepackage{color}
\usepackage{amsmath}
\usepackage{amssymb}
\usepackage{epstopdf}
\usepackage{verbatim}
\usepackage{breakcites}
\usepackage[version=3]{mhchem}

\newcommand{\be}{\begin{equation}}
\newcommand{\ee}{\end{equation}}

\begin{document}

\preprint{}

\title{A stochastic dynamical model of Arctic sea ice}

\author{W. Moon}
\affiliation{British Antarctic Survey \\ 
High Cross, Madingley Road, Cambridge, CB3 0ET, United Kingdom.}

\author{J. S. Wettlaufer}
\affiliation{Yale University, New Haven, USA}
\affiliation{Mathematical Institute, University of Oxford, Oxford, UK}
\affiliation{Nordita, Royal Institute of Technology and Stockholm University, Stockholm, Sweden}


\date{\today}

\begin{abstract}
The noise forcing underlying the variability in the Arctic ice cover has a wide range of principally unknown origins.  For this reason, the analytical and numerical solutions of a stochastic Arctic sea ice model are analyzed with both additive and multiplicative noise 
over a wide range of external heat-fluxes, $\Delta F_0$,  corresponding to greenhouse gas forcing.  
The stochastic variability fundamentally influences the nature of the deterministic steady state solutions corresponding to perennial, seasonal ice and ice free states.  Thus, the results are particularly relevant for the interpretation of the state of the system as the ice cover thins with $\Delta F_0$, allowing a thorough examination of the  differing effects of additive versus multiplicative noise.  In the perennial ice regime, the principal stochastic moments are calculated and compared to those determined from 
a stochastic perturbation theory described previously. As $\Delta F_0$ increases, the competing contributions to the variability 
of the destabilizing sea ice-albedo-feedback and the stabilizing long-wave radiative loss are examined in detail.  
At the end of summer the variability of the stochastic paths shows a clear maximum, which is due to the combination of the increasing influence of the albedo-feedback and an associated ``memory effect'', in which fluctuations accumulate from early spring to late summer.  This is counterbalanced by the stabilization of the ice cover due to the longwave loss of energy from the ice surface, 
which is enhanced during winter, thereby focusing the stochastic paths and decreasing the variability.  Finally, we discuss common examples in stochastic dynamics with multiplicative noise wherein the choice of the stochastic calculus (It\^{o} or Stratonovich) is not necessarily determinable {\em a-priori} from observations alone, which is why we treat both calculi on equal footing herein.
\end{abstract}


\maketitle

\section{\label{sec:intro} Introduction}
 
The advantages of simple deterministic theories of climate, such as clear assessment of stability and feedbacks, were evidently first recognized in the context of 
energy flux balance models independently by \citet{budyko1969} and \citet{sellers1969}.  Such approaches reveal key issues, such as the role of albedo feedback in planetary climate, the potential coexistence of multiple climate states under ostensibly the same forcing conditions, and the nature of the transition of mean states between them.  
Important early extensions of the original models including a form of meridional heat transfer are still analytically solvable, and can be used to assess the stability of high latitude ice-caps under varying climatic conditions \citep[e.g.,][]{held1974, north1975, North1981}. 
The inclusion of additional physics, such as diffusive type transport, can decrease the sensitivity of solutions relative to the simplest models \citep[e.g.,][]{Lindzen:1977} or bring out more stable solutions \citep[e.g.,][]{Rose:2009}, while sacrificing the ability to find analytical solutions.  Indeed, \citet{Lindzen:1977} point out that there is no {\em a-priori} compelling reason to assume that simple models with transport are superior than the Budyko-Sellers type of model.  In addition, solely deterministic models cannot capture the role of variability.  

In contrast, fully coupled climate models attempt to deterministically treat all of the processes in the climate system and to thereby capture the spatio-temporal structure of the atmosphere/land/ocean system.  Nonetheless, the inevitable complexity accompanying such treatments often precludes a clear identification of cause and effect in the absence of independent (e.g., observational) information.  However, this may be due to a confluence of real feedbacks and highly parameterized processes conspiring to obfuscate a variety of key interactions.  Moreover, in the Arctic projections vary widely among the IPCC models regarding the degree of ice loss through 2100 \citep[see e.g., Fig. 3 of][]{Eisenman:2011}. 

Stochastic climate models reside in a conceptual region between these two approaches having been introduced to develop a statistical understanding of the climate system or its subsystems
\citep[e.g.,][]{KH:76, North1981, benzi1981, nicolis1981, Saltzman:2002, DijkstraBook}.  In general the stochastic approach provides an important niche between solely deterministic low order models, which were not designed to treat high frequency variability, and complex fully coupled climate models.
In the spirit of the Langevin theory of Brownian motion, stochastic models typically consist of an underlying deterministic model augmented by stochastic forcing. The deterministic dynamics tends to embody the core physics of the system of note and the stochastic forcing captures the short-time scale processes which modify the deterministic dynamics.  Solutions of stochastic models provide the statistics underlying the variability that characterizes the interplay between the slow and fast dynamics.  This interplay introduces a complexity that can yield dynamics that are qualitatively different from simple deterministic models alone, while providing a richness that is seen in climate models, and yet still  within a framework amenable to analysis.   

The evolution of the air/sea/ice system has long been recognized as being a stochastic system \citep[e.g.,][and Refs. therein]{Lemke:1986}, and here we focus on a stochastic energy balance model of Arctic sea ice.  During the satellite era, in which we have high-fidelity measurements of the extent of the ice cover, there have been significant decreases in volume and extent \citep[see e.g.,][and refs therein]{OneWatt, Meier:2014}.   While both the observational record \citep{Agarwal2012} and climate model simulations \citep{Eisenman:2011} exhibit substantial variability on multiple time scales, it is clear that the mean minimum ice extent is decaying, which has stimulated the question of whether and when a seasonal ice state--no ice in the summer--may appear.  Importantly,  satellite data reveal that the nature of the noise itself is multifractal  \citep{Agarwal2012}, and thus given the prominence of variability in the observational record, a central question concerns how noise will impact the potential transitions in the state of the ice cover.  Because the observations show the complexity of the noise structure, and there is no a-priori evidence for a ``correct'' theoretical treatment (e.g., additive versus multiplicative) stochastic models must explore the influences of different, but rigorous treatments.

The response of the seasonal cycle of Arctic sea ice thickness to climate was first reproduced quantitatively in the thermodynamic model of \citet{maykut1971}.  The essence of this work has been captured more recently in several simpler models developed in the spirit of Budyko and Sellers to assess the question of the transitions between perennial, seasonal and ice free states \citep{thorndike1992, EW09}.  These approaches reproduce the observed season cycle of ice thickness and we use that of \citet{EW09} as the deterministic backbone of our stochastic model for the following reasons.  First, we have assessed in detail the stability 
of the deterministic steady states of this model and found the two key competing factors that dominate the response time scales \citep{MW:2011}.  In particular, the response time scales are governed by the destabilizing ice-albedo feedback and the stabilizing long-wave radiative energy loss which reflects the well known fact that thin ice grows more rapidly than thick ice \citep[see e.g., Fig. 2 of][]{MW:2011}.  Second, we have developed a perturbative framework of determining analytic solutions of the stochastic model that capture the key statistical moments of perennial ice states \citep{MW:2013}.  Third, the approach reveals a ``memory-effect'' whereby the intrinsic nonlinearity, asymmetry and stability characteristics of the interaction between the deterministic backbone and the noise provide an interpretive framework of cause and effect, along with their time scales.  Finally, numerical solutions to this model provide unique visualization of stochastic paths and probability density functions (PDFs) under the influence of increased greenhouse gas forcing ($\Delta F_0$).  This extends our analysis beyond the range available to our perturbative framework to allow examination of the dynamics of  seasonally-varying states. 

Because we can physically rationalize using both additive and multiplicative noise forcing on the same deterministic backbone, we present both here, although we note this makes for a rather weighty presentation. In particular, as discussed in detail in \S \ref {sec:model_numeric}.\ref{sec:num} below, in the case of multiplicative noise we give
the both stochastic calculi--It\^o and Stratonovich--equal weighting and thereby compare simulations using both.
 As we increase $\Delta F_0$ the stochastic stability of the system is examined in light of the expectations from the deterministic dynamics--transitions in the ice state are ``blurred'' by the variability in the stochastic paths.  The structure of the paper is as follows.  In the next section we describe the stochastic model and the numerical scheme.   We analyze the steady state stochastic solutions viz., stochastic paths, PDFs and 
statistical moments in \S \ref{sec:results}.  The overall dynamics is put in the framework of the ``ice-potential'', which is a seasonally evolving potential encoding the competition between stabilizing and destabilizing effects and how these change with $\Delta F_0$\footnote{The core dynamics are studied with the dimensionless version of the model but throughout this paper when we refer to values of $\Delta F_0$ they are understood to carry units of W m$^{-2}$}.  In this sense it is heuristically like an Ornstein-Uhlenbeck process, in a time-dependent potential, although we note that the deterministic backbone is nonlinear and nonautonomous.  We summarize and discuss the findings in \S \ref{sec:conclusion}.

\section{\label{sec:model_numeric}Stochastic sea ice model and numerical methods}

\subsection{\label{sec:model}Stochastic Arctic sea ice model}

 The stochastic Arctic sea ice model that forms the basis of our simulations has been described previously \citep[Eqs. 2 or 66 of][]{MW:2013}, but to 
 insure that this paper is
 self-contained we summarize it here.  The system is governed by a dimensionless Langevin equation written as 
 \begin{equation}
  dE=a(E,t)dt+b(E,t)\circ dW, 
  \label{eqn:sto_model}
 \end{equation}
 where the first term on the right side represents the deterministic backbone of the stochastic model, which is equivalent to that of \cite{EW09}, 
 and the second term treats the stochastic forcing where $dW$ represents a Wiener process, with $\circ$ denoting the Stratonovich interpretation of the noise as opposed to the It\^o interpretation, discussed below in \S \ref{sec:model_numeric} \ref{sec:num}. 
  
 The energy $E$ is defined as the amount of latent heat stored in a layer of ice of thickness $h$ or in the ocean mixed layer if the ice vanishes.  The convention used is that ice is present (absent) when $E$ is negative (positive).  The deterministic energy balance term $a(E,t)$ is
   \begin{align}
   a(E,t) &\equiv [1-\alpha(E)]F_S(t)-F_0(t)-F_T(t)T(t,E) \nonumber \\
   &+ \Delta F_0+F_B+\nu R(-E), \label{eqn:deterministic}
  \end{align}
   where
  \begin{align}
   \alpha(E) &= \frac{\alpha_{ml}+\alpha_i}{2}+\frac{\alpha_{ml}-\alpha_i}{2}\mbox{tanh}\left(\frac{E_0}{L_i h_{\alpha}}E\right) \label{eqn:albedo} {\text{and}}\\
   T(t,E) &= 
  \begin{cases}
   -\mathcal{R}\left[\frac{[1-\alpha(E)]F_S(t)-F_0(t)+\Delta F_0}{\frac{k_i L_i}{E~E^2_0}-F_T(t)}\right] & {\text{when~~}} E < 0 \\
   \frac{E_0}{c_{ml}H_{ml}}E & {\text{when~~}} E \geq 0.  
   \end{cases}
   \label{eqn:temp} 
  \end{align}
 Here, $\alpha(E)$ is the surface albedo and $T(t,E)$ the surface temperature.
The fraction, 1 - $\alpha(E)$, of the incident shortwave radiation $F_S(t)$ absorbed at the surface is modeled with an albedo function based on the Beer-Lambert law of exponential attenuation of radiative intensity with depth using a characteristic ice thickness $h_{\alpha} = 0.5m$ for extinction.  It captures the transition from perennial sea ice albedo ($\alpha_i$ = 0.68) to ocean albedo ($\alpha_{ml}$ = 0.2) and in this manner models the ice albedo feedback--it is clearly operative when the ice thickness approaches $h_{\alpha}$.  The core deterministic term $a(E,t)$ describes the energy flux balance at the atmosphere/ice (ocean) interface 
where we calculate the surface temperature $T(t,E)$. Quantitatively, this balance is dominated by incoming short-wave radiation, outgoing long-wave radiation and the conductive heat flux  through  sea ice.   During winter, the principal stabilizing mechanism is associated with how long-wave radiative loss drives ice growth -- thin ice grows more rapidly than thick ice \citep{Stefan}.  During summer, the principal destabilizing mechanism is the ice-albedo feedback, which becomes more prevalent when ice thins and approaches $h_{\alpha}$. The observed average annual export of $\sim$ 10\%  \citep{Kwok:2004} acts as a constant sink of energy, here represented by $\nu {\cal R}(-E)$, where $\nu=0.1$.  The ramp function is $\mathcal{R}(x\ge0) =x$ and $\mathcal{R}(x<0)= 0$, which captures the transition between freezing and melting states and the fact that sea ice export occurs only when sea ice is present.  

A more detailed description of the derivation of $a(E,t)$, including the incorporation of the various surface fluxes, the meridional heat flux due to large scale atmospheric  motions and the radiative transfer model is described in  \cite{EW09}. The stability of the deterministic model and the core competition between the destabilizing ice-albedo feedback and the stabilizing longwave loss at the ice surface is detailed in \citet{MW:2011}, which forms an important foundation for our interpretation of the stochastic dynamics. 

Recently, \cite{Wagner2015} suggested that the inclusion of a latitudinal variation in a deterministic single column model can substantially change the structure of the bifurcation diagram, thereby indicating that such complexities demarcate a model's ability to treat realistic behavior.  However, it is a basic result in the theory of dynamical systems \citep{Tredicce:2004} that, even in the simplest of models, when one constructs a bifurcation diagram with a slowly time-varying control parameter rather than a constant value, substantially different results are obtained. Hence, both complexity and the basic mathematical treatment, are important.

\subsection{\label{sec:albedo}The role of the sea-ice-albedo-feedback}

\begin{figure}[H]
\centering
\includegraphics[angle=0, scale=0.25,trim= 0mm 0mm 0mm 0mm, clip]{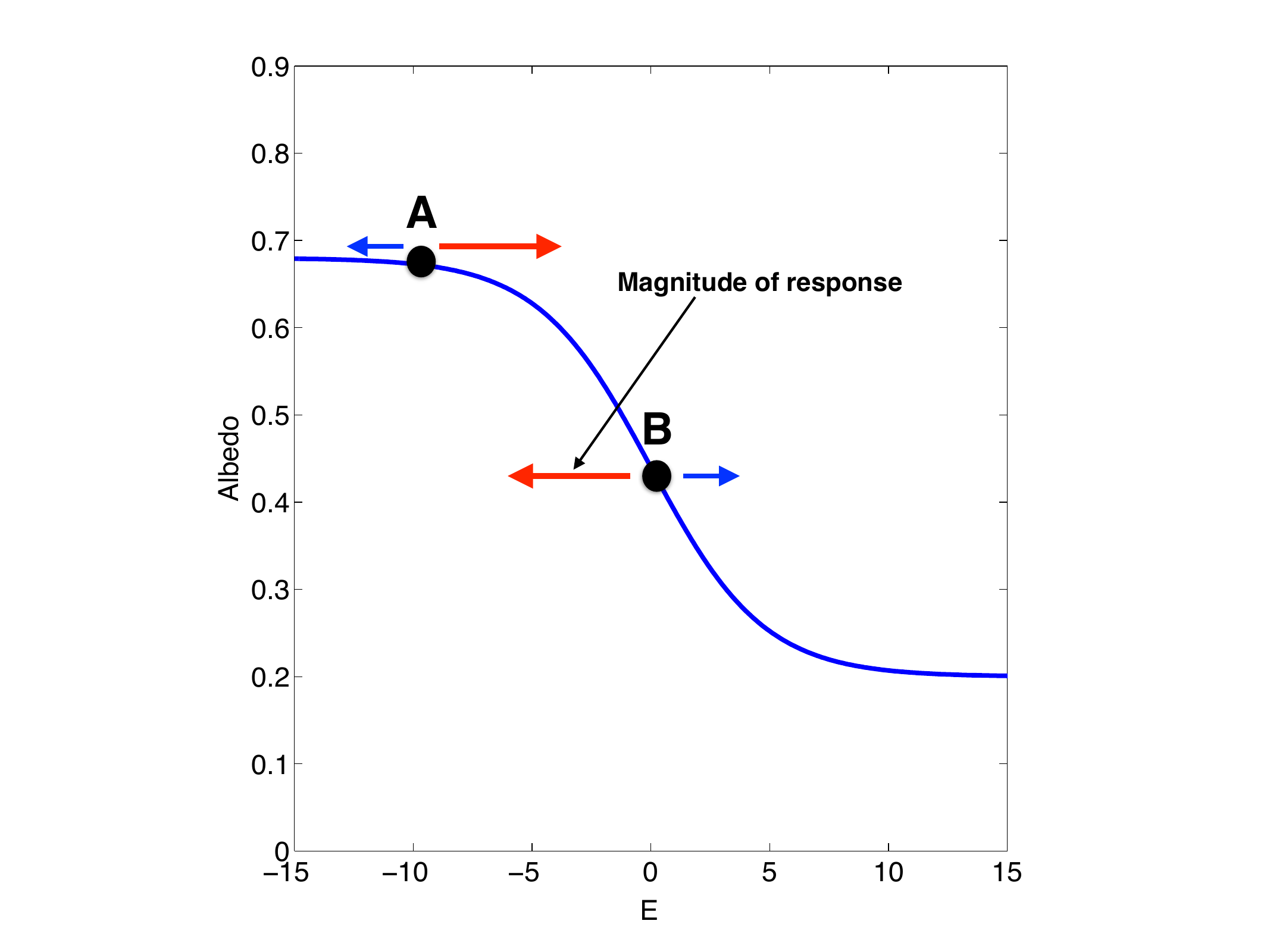}
\caption{Schematic diagram showing the relationship between magnitude of the response of sea-ice energy (thickness) and the albedo. 
         A and B represent two examples describing the asymmetric response of sea-ice to a given (signed) perturbation.
         The state A describes ice during summer when $\Delta F_0 \approx 19.0$, and the ice-albedo-feedback starts to operate.
         A positive perturbation will be more effective in changing the ice energy $E$ due to the sharp decrease of the albedo.
         The state B is relevant when $\Delta F_0 \approx 20.0$ and the sea ice is very thin during summer. Conversely to A, a negative
         perturbation will be more effective in changing the ice energy $E$ due to the sharp increase of the albedo.}
\label{fig:albedo_response}
\end{figure}

The most important process controlling the statistics of the stochastic solutions is the ice-albedo-feedback. The solution behavior is influenced by the asymmetric (signed) response of the ice to a perturbation associated with the dependence of albedo upon thickness as depicted in Fig.~\ref{fig:albedo_response}. 
The magnitude of the feedback depends upon the sensitivity of the albedo to a perturbation, which begins to become effective when $h \approx h_\alpha$.  The asymmetry is demonstrated for two ice states; A and B.  State A describes ice during summer when $\Delta F_0 \approx 19.0$.  Here, a positive (negative) perturbation will be more (less) effective in changing the ice energy $E$ due to the sharp decrease (small change) of the albedo.  State B describes ice during summer when $\Delta F_0 \approx 20.0$ and the sea ice is very thin. In contrast to A, a negative (positive) perturbation will be more (less) effective in changing the ice energy $E$ due to the sharp increase (small change) of the albedo.  Hence, very small changes in $\Delta F_0$ near this transition can generate highly variable stochastic paths.  This must be understood as a stochastic effect rather than a trend associated with increasing $\Delta F_0$; the key
point is that the variability increases with greenhouse-gas forcing.
This process is particularly important for understanding the solution statistics near the transition from perennial to seasonally varying ice-states.

\subsection{\label{sec:num}Numerical Method}

There are a wide variety of numerical methods used to solve stochastic ordinary differential equations \citep[e.g.,][]{kloeden1992}.
Most such methods rely upon a Taylor expansion, within either the It\^{o} or Stratonovich calculus framework. 
The order of numerical methods is determined by the convergence of either (a) the path of the solution itself--{\em strong convergence}, or (b) 
 the statistical moments--{\em weak convergence}.
The inclusion of one higher--order term in the Taylor expansion increases the numerical order by $0.5$ ($1.0$) in the sense of 
the strong convergence (weak convergence).
For many cases, order
$1.0$ ($2.0$) methods in strong (weak) convergence are sufficient. 
Here, we use a weak order $2.0$ approach based upon the Runge-Kutta method of \cite{tocino2002}. The 
discrete form of the Eq. (\ref{eqn:sto_model}) is written as
\begin{align}
 E_{n+1}=&E_n+\frac{1}{2}(k_0+k_1)\Delta+\frac{1}{4}(2s_0+s_1+s_2)\Delta W_n \nonumber \\
 &+\frac{1}{4}(s_2-s_1)
 \left(\sqrt{(\Delta)}-\frac{(\Delta W_n)^2}{\sqrt{(\Delta)}}\right),
\end{align}
where
\begin{align}
&k_0 = \overline{a}(E_n, t_n), \\
&s_0 = b(E_n,t_n), \\
&k_1 = \overline{a}(E_n+k_0\Delta+\Delta W_n s_0,t_n+\Delta), \\
&s_1 = b(E_n+k_0\Delta +\sqrt{\Delta}s_0,t_n+\Delta), \text{and} \\
&s_2 = b(E_n+k_0\Delta-\sqrt{\Delta}s_0,t_n+\Delta).
\end{align}
If the stochastic model is interpreted within the framework of It\^{o}  calculus then $\overline{a}(E,t) \equiv a(E,t)$, whereas in the framework of Stratonovich calculus, 
$\overline{a}(E,t) \equiv a(E,t)+\frac{1}{2}b(E,t)\partial_E b(E,t)$. This transformation between the two forms of stochastic calculus was introduced by \cite{WZ65}, and there are
more pedagogical discussions of the mathematical background and geophysical applications found in \cite{GFD} and \cite{DijkstraBook}. 
The time step is $\Delta$ and 
$\Delta W_n$ is a Gaussian variable whose mean and standard deviation are $0$ and $\sqrt{\Delta}$, respectively.
Using this method and converting to dimensional time the system reaches a steady state in 20 years.  
To generate ensemble statistics, we repeat the simulation using different values of $\Delta W_n$ and 
different values of $\Delta F_0$.  The baseline numerical analysis uses $10^6$ ensemble simulations with a $10^{-6}yr$ time step and a noise intensity of $0.05$. 

\subsubsection{\label{sec:mult}Additive and Multiplicative noise structure}

Clearly, the simplest form of additive noise transforms the function $b(E,t)$ on the right hand side of Eq. (\ref{eqn:sto_model}) into a (typically small) constant $b$.  This is generally referred to as constant additive noise.  However, since we are dealing with a deterministic dynamics that is time-periodic, there are a range of possible additive noise scenarios that can be treated.  We describe our approach presently.  

We introduce multiplicative noise through variability in sea ice export, which we can ascribe to the observation that the geostrophic wind field that drives ice motion can be treated as a Gaussian random field \citep{ThorndikeJAS:1982, Agarwal2017}.  
To include the effect of fluctuations upon the sea ice export, we introduce a random variable as
$\nu = \nu_0+\sigma\xi(t)$, where the constant value $\nu_0=0.1$ becomes that from the deterministic dynamics and $\xi(t)$ is related to the Weiner process as $\xi(t) = dW/dt$.  Hence, 
$b(E,t)$ of Eq. (\ref{eqn:sto_model}) becomes $\sigma \mathcal{R}(-E)$ and we can rewrite the 
stochastic model as
\begin{equation}
  dE=a(E,t)dt+\sigma \mathcal{R}(-E)\circ dW, 
 \label{eq:fullmodel}
\end{equation}
where the noise amplitude $\sigma$ is small relative to unity (for our numerical studies it is set to $0.05$), and  
the deterministic term $a(E,t)$ is as in Eq. (\ref{eqn:deterministic}), but with $\nu \rightarrow \nu_0 = 0.1$. 
%

It is important to note that even in well studied nonlinear systems, the mathematical {\em and} physical interpretation of multiplicative noise depends upon the choice of stochastic calculus and there are subtle issues arising even in the simplest form of additive noise. 
A core difference between the calculi resides in the freedom to choose the value of the integrand in a subinterval of the Riemann sum.  
For example, It\^{o} calculus is often preferred because it preserves the Martingale property, wherein the expectation value of any time-dependent quantity
depends solely upon the present value. Although this approach has many practical numerical advantages, the usual rules of calculus are not obeyed, whereas this is not the case with Stratonovich calculus.  In this setting, the major difference between Stratonovich and It\^{o} calculus is the shift of the mean value due to the accumulation of noise-forcing. 
Here, we will consider both perspectives numerically through simultaneous treatment of 
the statistical moments and the stochastic paths.

\cite{WZ65} argued that there is no real world system in which perfect white noise exists.  Thus, Brownian motion $x(t)$ 
approximates a description $x_n(t)$ that is continuous with at least a piece-wise continuous derivative.  
By showing that $x_n(t) \rightarrow x(t)$ as $n \rightarrow \infty$ they recovered Stratonovich calculus.  
Accordingly, the choice of stochastic calculus resides in the characteristics of the noise and continuity arguments \citep{MW:2014}. 
On one hand, in statistical physics white noise is typically defined through a $\delta-$autocorrelation, and it is also suggested that this definition is equivalent to Stratonovich calculus \citep{Risken1996}.  Thus, the use of white noise to approximate high frequency processes in systems observed over much longer time scales is often argued to be within the purview of Stratonovich calculus. 
On the other hand, in finance and biology, where most of the high frequency processes are {\em assumed} to be discrete, and hence
the above arguments may not be applied.  Thus, It\^{o} calculus is {\em assumed} to be appropriate \citep{Turelli1977, Shreve2004}, 
thereby maintaining the Martingale property. In terms of overall separation of time scales, there is no conceptual distinction between the statistics of  
water molecules colliding pollen grains and the trading equities (or the like).  
Hence, the question remains if, how and when it is appropriate to use continuity considerations 
as a core criterion to choose either of the calculi being discussed here. 

We believe the choice of which stochastic calculus should used for a particular set of physical processes is more complicated than the above. 
For example, in building a mathematical model 
it is common to ignore the influence of high-frequency processes on the deterministic dynamics, 
although we know there are situations when this is a poor assumption, 
such as in the presence of inertial \citep{Kupferman2004} or feedback \citep{Pesce2013} effects. 
Indeed, when Kupferman and colleagues \citep{Kupferman2004} studied systems
with multiplicative colored noise and inertia they found that if the correlation time of the noise is faster (slower) than the relaxation time, 
this leads to the It\^{o} (Stratonovich)
calculus form of the limiting stochastic differential equation. 
Similarly, It\^{o} calculus is invoked to interpret experiments wherein the time delay of the feedback is much
larger than the noise correlation time \citep{Pesce2013}. Hence, there is an experimental demonstration that the choice of the stochastic calculus 
is not necessarily {\em a-priori} determinable from observations alone.  Indeed, even taking the white noise limit of a colored noise process, which leads to It\^{o}
calculus, this is a deliberate choice, which is often made for numerical reasons; principally the appeal of the aesthetics of the standard forward Euler scheme \citep{GFD}.  

For these reasons, and those found in a more detailed discussion \citep{MW:2014}, we take an agnostic approach and treat the two calculi with equal weighting. Although the noise structure of Arctic sea ice can be quantified  from observations \citep{Agarwal2012}, the choice of stochastic calculus cannot be deduced from them. 
Hence, we believe that comparing the stochastic solutions from the two calculi will be beneficial to those trying to implement 
stochastic models in a variety of contexts.


For the most general multiplicative noise case, $\sigma \mathcal{R}(-E(t))\xi(t)$, with $|\sigma|\ll1$, we have a theoretical framework with which to compare our numerical results \citep{MW:2013}.  Because the theoretical framework is perturbative, several more cases are then naturally structured for comparison.  We define seasonally-varying additive noise (SVA) when the noise-magnitude is $\sigma \mathcal{R}(-E_S(t))$, where $E_S(t)$ is the deterministic steady-state solution and hence the noise is additive, but time varying with the seasonal cycle.
The constant additive noise case (CA) is a natural limit of the seasonally-varying additive noise case and has noise amplitude $\sigma\overline{\mathcal{R}(-E_S(t))}$, where the overbar is the seasonal time average of $E_S(t)$.  The difference between these two cases reveals the impact of seasonally varying noise-magnitude.  

It is prudent to deal with all of these cases because the first-order perturbative solution is equivalent to that with seasonally varying noise, and the effect of multiplicative noise upon the steady-state stochastic solutions does not appear until second order.   Therefore, we will compare the full model described by Eq.~(\ref{eq:fullmodel}) with the seasonally varying noise case to reveal the bare effect of multiplicative noise.

Four similar but systematically different cases will be analyzed and compared. We first compare the constant additive noise (CA) and the seasonally varying additive 
noise (SVA) cases and then the two different stochastic calculi; It\^{o} (IM) and Stratonovich (SM), where the M denotes multiplicative. 
This allows us to compare and contrast the role of different classes of noise-forcing.

\section{\label{sec:results}Results : Additive Noise Cases}

In this section we describe a large suite of simulations of this model using the numerical method explained in the previous section.  We obtain many stochastic realizations
and generate ensemble statistics. As noted above, the amplitude of the constant additive noise-forcing is fixed at $0.05$ for all simulations. One advantage of the numerical
simulations over our solely analytical method is the ability to observe the evolution of a specific stochastic realization and to directly construct a probability 
density function (PDF) for a given type of noise forcing. Moreover, we can explore the stochastic solutions over a range of $\Delta F_0$ in which our stochastic
perturbation method cannot be applied, although we will still compare the numerical solutions to the analytic solutions over their range of validity \citep{MW:2013}. We thereby extend our understanding and analysis to the seasonally varying states, where stochastic effects are particularly important.

Depending on the geometric structure of the deterministic backbone of the model in the vicinity of the steady-state solution,  
and the nature of the noise-forcing, the stochastic solution will exhibit dispersion 
relative to the deterministic solution, giving rise to asymmetry in stochastic realizations.


\begin{figure}[H]
\centering
\includegraphics[angle=0,scale=0.25,trim= 0mm 0mm 0mm 0mm, clip]{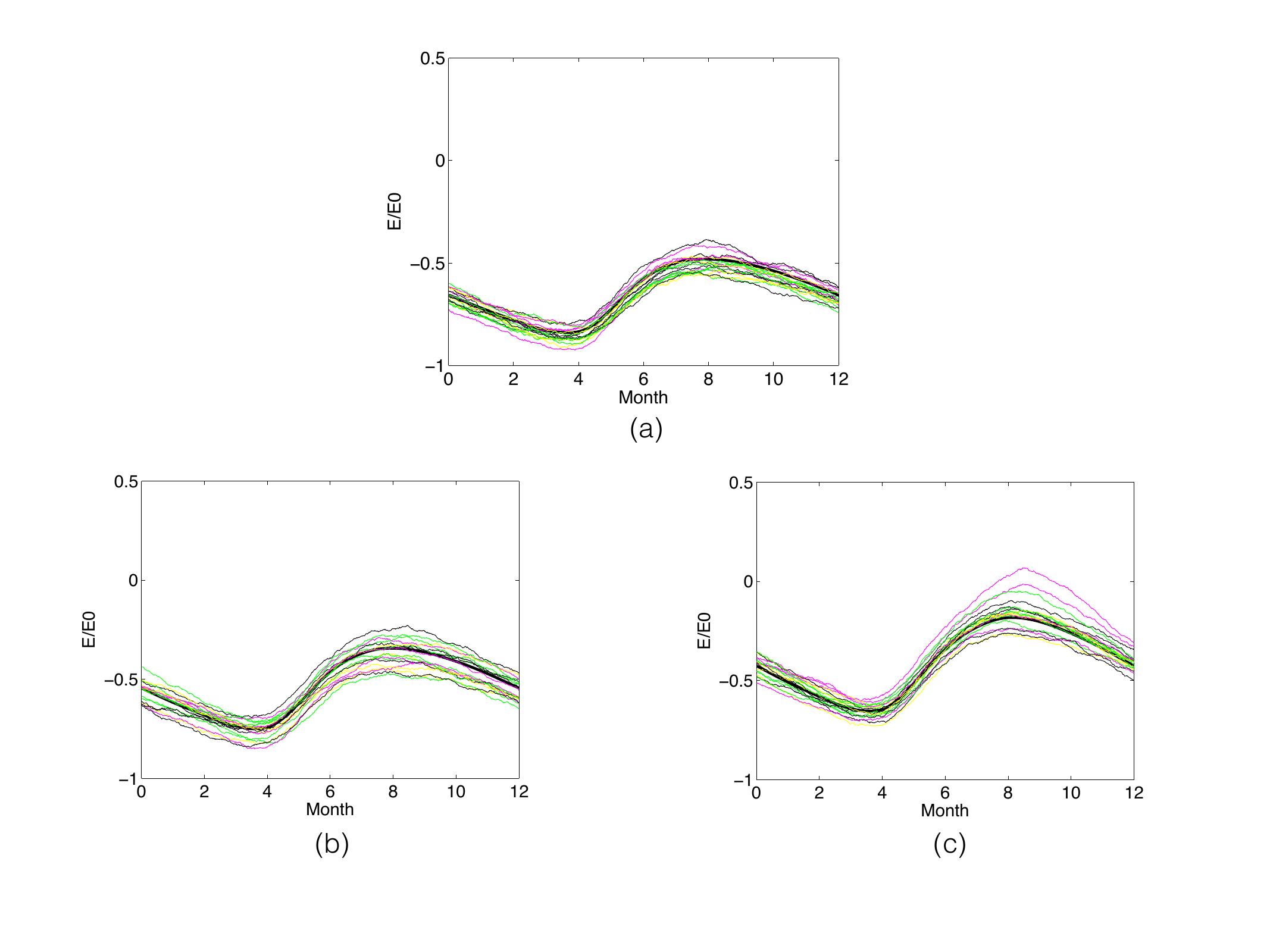}
\caption{Several realizations of the seasonal cycle of the stochastic solutions with three different values of $\Delta F_0$; (a) $10.0$, (b) $14.0$, (c) $18.0$.
         The thick black lines represent deterministic stable seasonal cycles of sea-ice-thickness. The other lines show 
         different realizations of the stochastic solutions.}
\label{fig:sample_path}
\end{figure} 

\subsection{Perennial ice-states}

Deterministic perennial ice-states exist under greenhouse-gas forcing $\Delta F_0$ up to $\sim 20$ when a continuous
transition to a seasonally varying state occurs, whereas beyond $\sim 23$ the seasonal ice vanishes in a saddle node bifurcation to a perennial ice free state \cite[see Fig. 3 of][]{EW09}. Thus, as a first example, in Figure~\ref{fig:sample_path} we show stochastic realizations as $\Delta F_0$ grows from $10$ to
$18$. It is noticeable that even with the same noise-forcing,
the spread of the stochastic realizations increases with $\Delta F_0$, with some realizations exhibiting seasonal ice states 
under forcing in which the deterministic state has perennial ice.

As described previously \citep{MW:2013}, the stochastic model can be represented in an approximate form near the deterministic solution $E_S(t,\Delta F_0)$ as follows. If
we let $E(t)=E_S(t)+\eta$, where $\eta$ is 
the departure from the deterministic solutions
and is written as
\begin{equation} \label{eqn:sto_eq_ch06}
 \frac{d\eta}{dt}=c(t)\eta+d(t)\eta^2 + \sigma \xi,
\end{equation}
where $c(t) \equiv \frac{\partial a(E,t)}{\partial E}|_{E=E_S}$ and $d(t) \equiv \frac{1}{2}\frac{\partial^2 a(E,t)}{\partial E^2}|_{E=E_S}$, and $a(E,t)$ is that from Eq. \ref{eqn:sto_model}. 
Here, we introduce
the ``ice potential'' $V(\eta,t) \equiv -\frac{1}{2}c(t)\eta^2-\frac{1}{3}d(t)\eta^3$ to rewrite Eq.~\ref{eqn:sto_eq_ch06} as
\begin{align}
 \frac{d\eta}{dt}=-\frac{\partial}{\partial \eta}V(\eta,t)+\sigma\xi.
\end{align}
The interpretation of both the analytic and numerical solutions is facilitated by examining the structure of the potential $V(\eta,t)$, which 
reflects the geometry of the deterministic solutions\footnote{Note, because $\eta$ is an energy variable we discuss $V(\eta,t)$ and $V(E,t)$
interchangeably}.

\begin{figure}[H]
\centering
\includegraphics[angle=0,scale=0.25,trim= 0mm 0mm 0mm 0mm, clip]{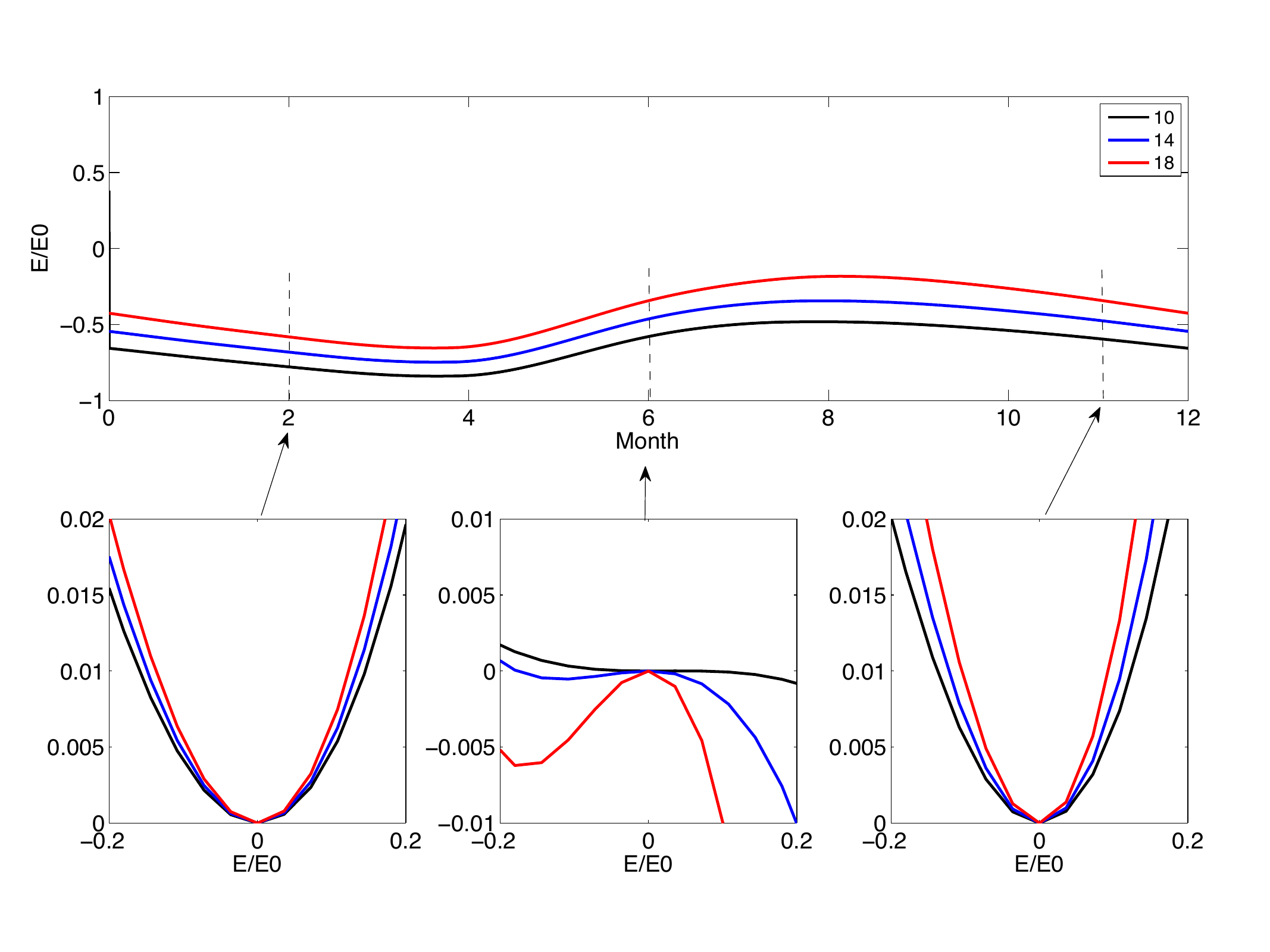}
\caption{The upper panel shows the stable periodic steady-state solutions at three different values of $\Delta F_0$; $10.0$ (black), $14.0$ (blue) and $18.0$ (red), and the lower panel shows 
the potential $V(E,t)$ for the same values of  $\Delta F_0$ in February, June and November, respectively.  The sign of $E/E_0$ is the same as the sign of $\eta$.}
\label{fig:potential1}
\end{figure}

In an autonomous dynamical system, only a single potential controls the influence of a given perturbation. Here, we have a periodic nonlinear non-autonomous 
dynamical system, which is much more complicated because the potential evolves continuously. The instantaneous stability of the 
system is reflected in the shape of $V$ (concave or convex).  However, as shown in Fig.~\ref{fig:potential1}, the potentials are not symmetric about the deterministic solutions. 
The response of the system to a perturbation is dependent on its sign and is proportional to the slope of $V$.  
This is understood as being due to 
the nonlinearity in $a(E,t)$, which is reflected in $d(t)$. The essence of the nonlinearity is that at a given time the response time-scale is dependent upon the state of the system--the sea ice thickness.

The potentials during the cold periods shown (February and November) are concave (Fig.~\ref{fig:potential1}). 
As $\Delta F_0$ increases the concave minima deepen (compare for example $\Delta F_0 = 10.0$ and $14.0$).
Physically, this reflects the long understood phenomenon that thinner ice grows faster than thicker ice \citep{Stefan}. 
Heat conduction is proportional to $\frac{\Delta T}{h}$, where $\Delta T$ is the temperature difference between the top and the bottom of sea ice
of thickness $h$. Because the growth-rate of the ice depends on how efficiently the latent heat and the oceanic heat-flux can be conducted through it,
for the same surface-heat-balance thin ice grows faster than thick ice.
In contrast, near zero the potentials during the summer are convex, and the 
asymmetry about the origin becomes larger as $\Delta F_0$ increases. For $\Delta F_0 = 10.0$, the potentials in June or July are almost flat near the origin. 
However, when $\Delta F_0 = 18.0$, the potentials at the same time are convex 
and asymmetric with the magnitude of the slope being larger for $\eta > 0$ ($E/E_0 > 0$). 
The origin of this behavior is that the ice-albedo-feedback is more sensitive as the ice thins and the magnitude of the energy $|E|$ decreases. 

The potentials shown in Fig.~\ref{fig:potential1} demonstrate
the overall seasonal variation. The potentials for November and February reflect the longwave stabilization during winter,
suppressing the effects of perturbations, and those for June show that the effect of the ice-albedo-feedback is to amplify the magnitude of a 
perturbation. These two main processes combine with the effects of stochastic forcing determine the steady-state stochastic solutions of the model.  We stress that although we show several examples of $V(\eta,t)$ in Fig.~\ref{fig:potential1}, the potential changes continuously in time thereby impacting the stochastic paths. 

\begin{figure}[H]
\centering
\includegraphics[angle=0,scale=0.25,trim= 0mm 0mm 0mm 0mm, clip]{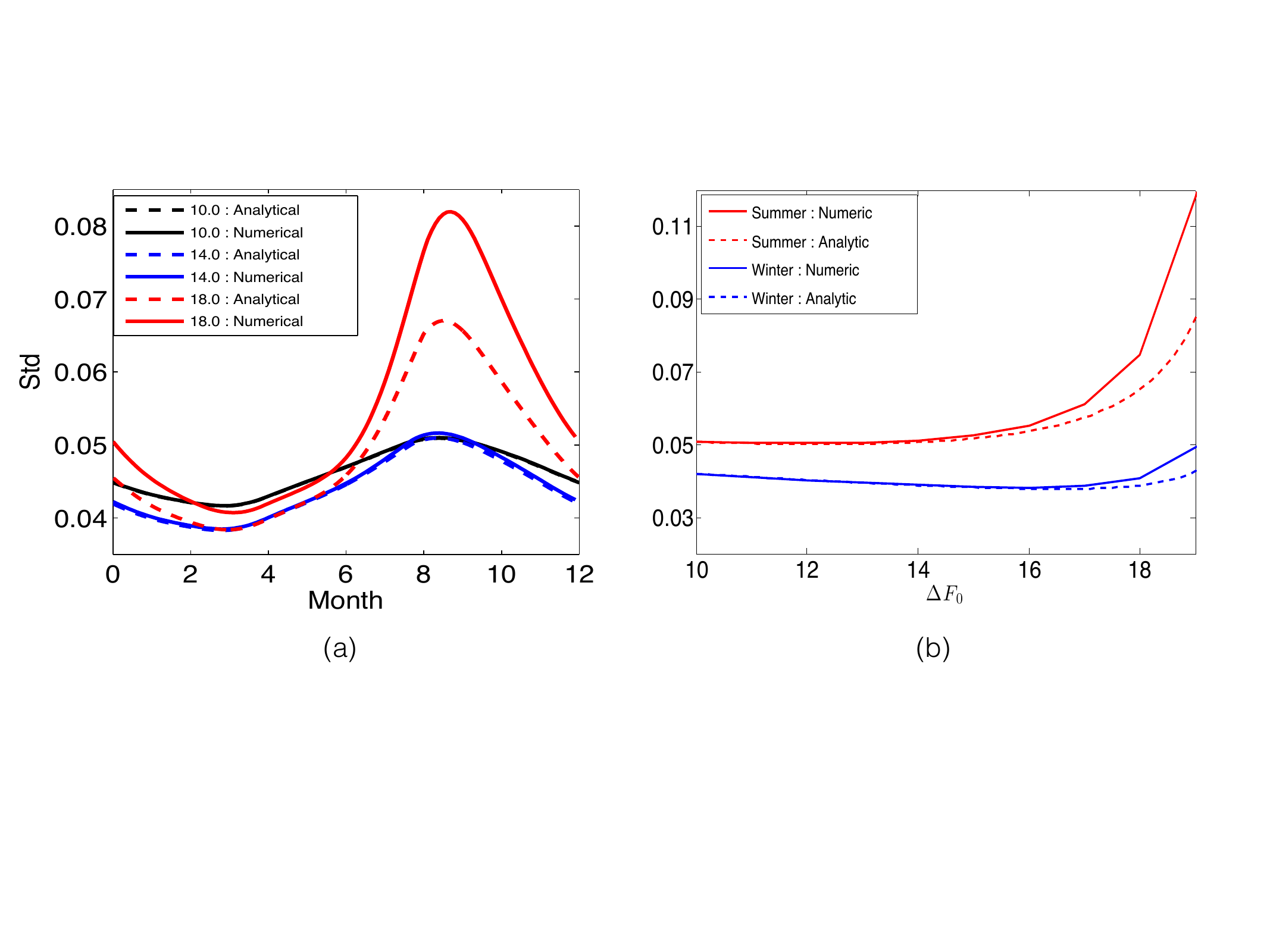}
\caption{Comparison between analytic (dashed) and numerical (solid) solutions for (a) the standard deviation of the seasonal cycle for the
         three different values of $\Delta F_0$ shown in the legend and (b) and as a function of $\Delta F_0$ (b).}
\label{fig:comparison_perennial_seasonal}
\end{figure}

 As we have described previously \citep{MW:2013}, the steady-state stochastic solutions are determined by the cumulative influence of the potentials in the time domain,  which is scaled by the response time of the deterministic solutions. This rectification was referred to as the ``memory effect''.   The stochastic paths change continuously as the potential  $V(\eta,t)$, changes, exhibiting a clear seasonality of trajectories.  At the end of the winter (summer), the stochastic paths are more concentrated (widely distributed) about  $E/E_0 = 0$, reflecting the deterministic physics of longwave radiative stabilization and the destabilizing ice-albedo feedback.  There is little difference between $\Delta F_0 = 10$ and $14.0$, but as $\Delta F_0$ increases to $18.0$, the stochastic paths are more widely distributed, as can be seen in the Supplementary Material.  In particular, the variability of the paths at the end of summer exhibit a clear maximum (Fig.~\ref{fig:comparison_perennial_seasonal}), which is due to the combination of the increasing importance of the ice-albedo-feedback and the associated memory effect accumulating a signal from early spring to late summer.  

The seasonality of the PDFs can be understood in terms of the memory effect. For example, the PDFs in March 
have sharp peak near the deterministic steady-state solution ($E/E_0=0$). This is explained by the concave shape of the potentials from 
September to February, which insures that perturbations converge to $E/E_0=0$. 
Conversely, the destabilizing  effect of the ice-albedo-feedback is cumulative, beginning in April or May and reaching a maximum by
the end of summer.
At any instant the stochastic solutions embody the delayed effect of these competing destabilizing and stabilizing processes. For example, while the ice-albedo-feedback begins in April or May and is active all summer, the PDFs are not significantly positively skewed until the end of summer.   This reflects the memory effect.


\subsubsection{Validity of Analytical Solutions}
Previously we calculated perturbatively the first three moments using (in part) the noise amplitude $\sigma$ as a small parameter, finding the standard deviation at $O(\sigma)$, and the mean and 
skewness at $O(\sigma^2)$ \citep{MW:2013}.
In Fig.~\ref{fig:comparison_perennial_seasonal}, we compare the analytic solutions with the numerical solutions 
for three different values of 
$\Delta F_0$.   The match between two 
solutions is excellent for lower values of $\Delta F_0$, but, as expected, the deviation grows with $\Delta F_0$.
The deviation between the theory and the numerical solutions that starts to appear as $\Delta F_0$ increases is due to fact that, at first order, the perturbative solutions fail to include the asymmetric effects associated with the ice-albedo-feedback as seen in the structure of the seasonally varying potentials. 

The essence of the perturbative theory is that to first approximation the PDF is Gaussian with a mean equal to
that of the deterministic steady-state solutions and the standard deviation changes periodically depending on the time-dependent state of stability as reflected in the ice-potential.
The deviation of the stochastic means from the deterministic solutions and the skewness appear at second order.
The basic behavior of the solutions at each order is determined by the interplay between the stability of the ice, the nonlinearly induced asymmetry in the response, and the intensity of the noise-forcing.
In particular, the analytic solutions nicely describe the memory effect in the form of a delayed integral, which is used to interpret the seasonality of the stochastic 
solutions.
The memory effect combines the cumulative influence of the interaction between the statistical fluctuations over the seasonal cycle and the stabilizing and destabilizing processes embodied in the deterministic ice-potential, which is also reflected in the Floquet exponents of the deterministic solutions.

The increasingly non-Gaussian behavior as $\Delta F_0$ increases demonstrates the limitations of the analytic method. The reason
for this deviation is clear; the method is based upon small-amplitude noise-forcing and thus implicitly assumes that the behavior of stochastic
paths is mainly controlled by the stability and the asymmetry embodied in the deterministic solutions. Such behavior depends principally 
upon the characteristics inherited from the deterministic solutions, rather than the stochastic paths. 
However, when $\Delta F_0$ is large, thin ice is particularly sensitive to the ice-albedo-feedback. Therefore, the stochastic
paths are not only affected by the stability and the asymmetry of the deterministic dynamics but they are also highly dependent upon the noise induced variability. 
For example, positive stochastic forcing during summer is magnified due to the ice-albedo-feedback and then 
significantly damped during winter by the intensification of the longwave stabilization. This leads to a larger response of the statistical moments 
relative to the analytic solutions. 

Moving out of the range of validity of the analytical framework, in the next section we will study the regime of $\Delta F_0$ where we have stable seasonally varying states.  However, we can still rely on the theory to interpret solutions within the context of the behavior of the local ice potentials.

 \subsection{Seasonal ice states} 
 
According to the deterministic theory, the transition from a perennial ice state to 
a seasonally varying state (with an ice-free summer) is continuous and reversible as $\Delta F_0$ increases \citep{EW09}. 
Approaching this transition, when still in the perennial state, the response time-scale of a perturbation to the deterministic dynamics is approximately 5-years.  However, once the stable seasonally varying state emerges, the response time scale abruptly drops to 2-years \citep{MW:2011}. From the perspective of a stochastic model, this transition is far less clear because noise-forcing acts as an additional heat-flux source or sink. Intuitively, this implies that the two states can statistically coexist 
with the same $\Delta F_0$, thereby generating a great deal of variability relative to that of states deeply in the perennial ice regime.  As $\Delta F_0$ further increases, the deterministic system approaches a saddle-node bifurcation from a seasonally varying state to an ice-free state  \citep{EW09}.
It is important to investigate the variability of these states near the bifurcation point. In this section we study the entire range of $\Delta F_0$ spanning these transitions.

 \subsubsection{The transition from perennial to seasonal ice}
 
The seasonal state appears in the deterministic dynamics as $\Delta F_0$ approaches $20.5$ from below.  Now, we investigate
the characteristics of the stochastic solutions near this transition, which is ``blurred'' in the sense that two stable states coexist at a single $\Delta F_0$. 
\begin{figure}[H]
\centering
\includegraphics[angle=0,scale=0.25,trim= 0mm 0mm 0mm 0mm, clip]{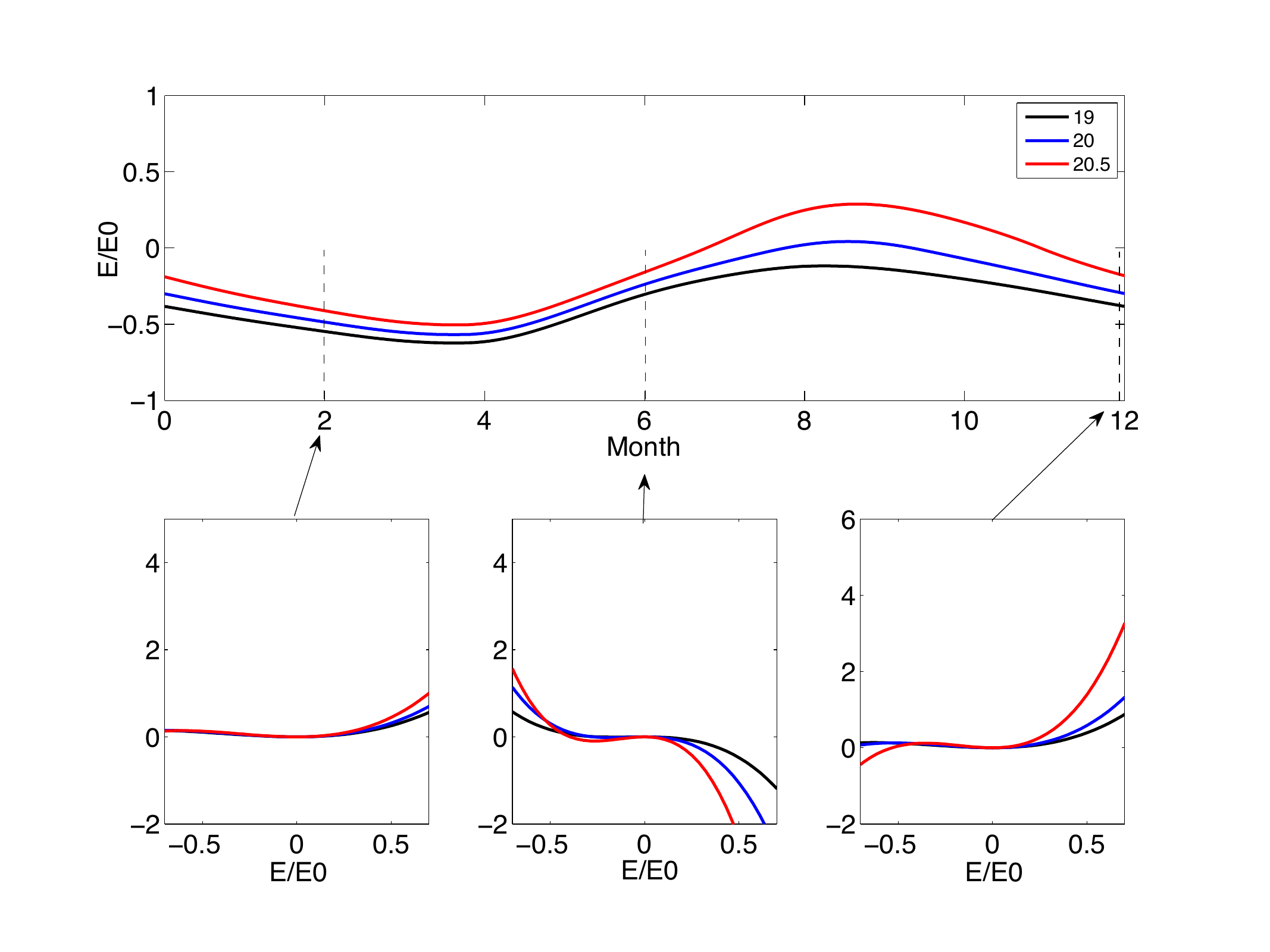}
\caption{Same as figure~\ref{fig:potential1}, but here the three different values of $\Delta F_0$ are $19.0$ (black), $20.0$ (blue) and $20.5$ (red).  
         The potentials are shown in February, June and December.}
\label{fig:potential2}
\end{figure}

\begin{figure}[H]
\centering
\includegraphics[angle=0,scale=0.25,trim= 0mm 0mm 0mm 0mm, clip]{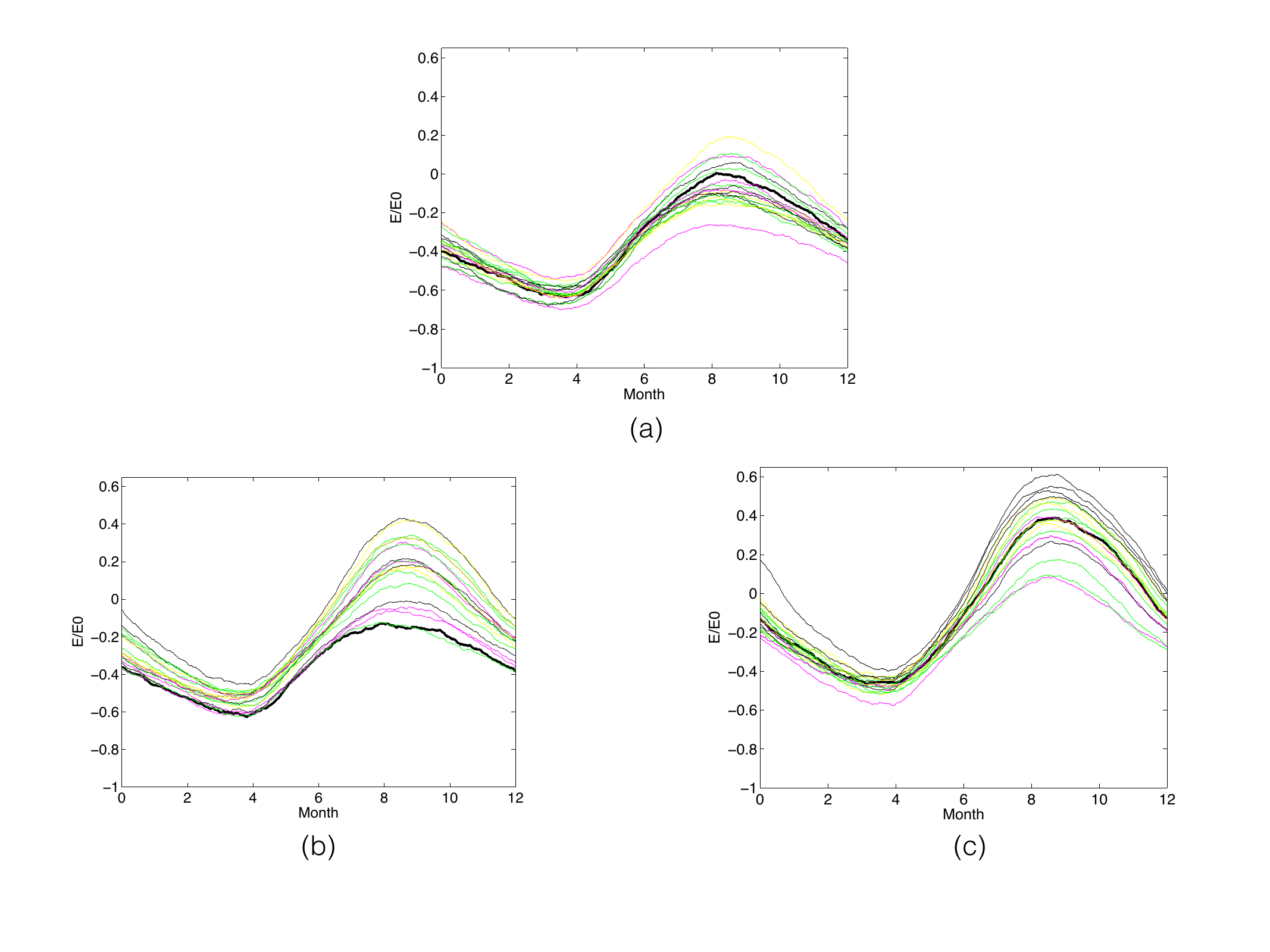}
\caption{Several realizations of the seasonal cycle of the stochastic solutions with three different values of $\Delta F_0$; (a) $19.0$, (b) $20.0$, (c) $20.5$.
         The thick black lines represent deterministic stable seasonal cycles of sea-ice-thickness. The other lines show 
         different realizations of the stochastic solutions.}
\label{fig:sample_path2}
\end{figure} 

We derive intuition by examining the ice potential $V(E,t)$ near the steady-state solutions and in Fig.~\ref{fig:potential2} we plot potentials for February, June and December when $\Delta F_0$ is $19.0$, $20.0$ and $20.5$.  Because the deterministic steady-state 
solutions contain very thin sea ice or open ocean during summer, we see enhanced competition between the ice-albedo-feedback and the longwave
stabilization, and hence the asymmetric response of the system to a given perturbation. It is instructive to focus on the potentials for $\Delta F_0=20.5$. The potential for June exhibits the ice-albedo-feedback through the strong negative slopes when $E/E_0 > 0$. 
A positive perturbation will grow rapidly away from the steady-state solution; for example, melting leads to more melting due to an additional decrease of 
the ice-albedo. By parity of reasoning a negative perturbation leads to more ice -- the albedo feedback is always positive.
However, as we have an energy balance model based on heat conduction, and the albedo treatment is based on radiative extinction, the albedo feedback becomes
strongly operative once the ice thickness $h\simeq h_{\alpha}=0.5m$. 
This ice thickness enhancement of the asymmetric sensitivity
induced by the ice-albedo-feedback is fruitfully demonstrated by examining the detailed changes in the ice potentials. 

Recall from Fig.~\ref{fig:potential1} that during winter when $\Delta F_0$ is such that the system in the perennial ice-state, the potentials are concave and longwave 
radiative loss strongly stabilizes perturbations in a symmetric manner. However, when $\Delta F_0$ increases and the ice is thinner, Fig.~\ref{fig:potential2} shows that during winter the longwave stabilizing response to a perturbation is highly asymmetric.  Clearly the slope on the positive side is much larger than that on the negative side and this asymmetry increases with $\Delta F_0$. 

In Fig.~\ref{fig:sample_path2} we see stochastic realizations as $\Delta F_0$ transitions from perennial to seasonal ice-states.  (These are discussed in terms of the comparison between additive versus multiplicative noise in more detail in the Supplementary Material and in \S \ref{sec:resultsmult}.)  Although the deterministic steady-state solutions for $\Delta F_0 =19.0$ and $20.0$ are still perennial ice-states, the ice is quite thin during the summer and the stochastic realizations tend toward seasonally varying states with ice free summers. 
Moreover, while the longwave stabilization is stronger for thinner ice, the ice-albedo-feedback dominates, and the asymmetry associated with the latter is stronger than that associated with the former.   Recall that as the ice thickness approaches $h_\alpha$ the ice albedo changes from that of perennial ice (0.68) to that of open ocean (0.2).  Hence, depending on whether the ice thickness is large or small relative to $h_\alpha$ the response to a perturbation will be very different.  Namely, when $h \approx h_\alpha$ the ice is more sensitive to a positive (negative) perturbation which causes a dramatic increase (decrease) in the albedo.  For this reason, near the transition from the perennial to the seasonal ice state, the summer ice-thickness approaches $h_\alpha$ and a new asymmetry in the stochastic ensemble statistics emerges.  Interestingly, we then find that as the system approaches the deterministic transition to seasonal ice, the ice-albedo-feedback drives the stochastic solutions towards the seasonal state.  However, with only a small increase in $\Delta F_0$, the stochastic solutions tend towards the perennial state.   This suggests that near the deterministic transition to seasonal ice, the statistical fluctuations in the ice cover can exhibit behavior of both states and thus the transition itself cannot be explained using concepts based on linear response.

\subsubsection{Approaching the deterministic saddle-node bifurcation}

\begin{figure}[H]
\centering
\includegraphics[angle=0,scale=0.25,trim= 0mm 0mm 0mm 0mm, clip]{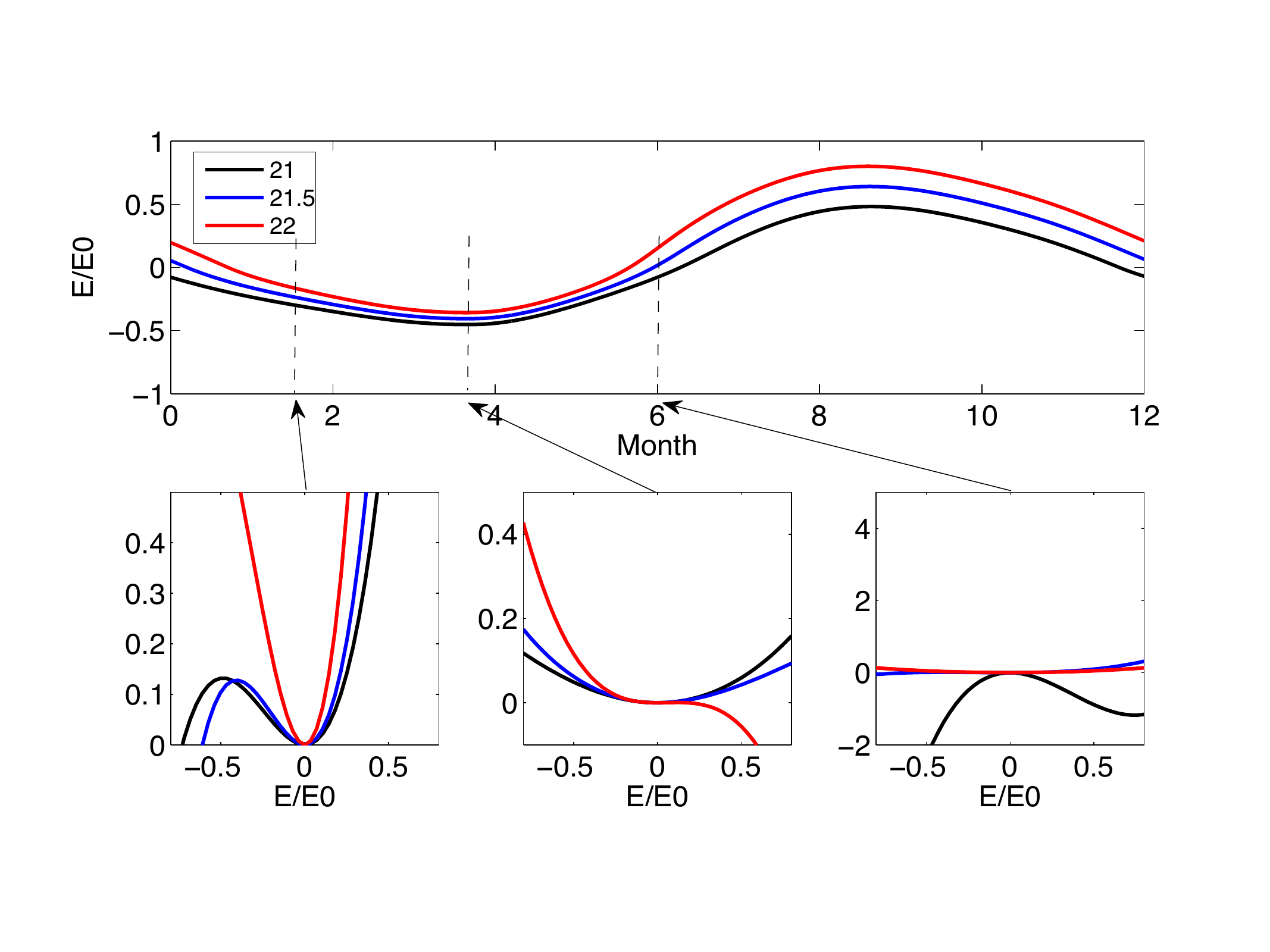}
\caption{Same as figure~\ref{fig:potential1} but for three different values of $\Delta F_0$ of $21.0$, $21.5$ and $22.0$ as shown in the legend. 
         The three months indicated in the lower panels are January, March and May.}
\label{fig:potential3}
\end{figure}

As $\Delta F_0$ increases, the deterministic seasonally varying ice states approach a saddle-node bifurcation to an ice-free state ($\Delta F_0$ = 23), which is separated from the perennial state by a hysteresis loop.  Here again to examine the stochastic solutions we consider the seasonal cycle of the potentials $V(E,t)$ of the deterministic steady-state solutions 
 for $\Delta F_0 = 21.0$, $21.5$ and $22.0$ (Fig.~\ref{fig:potential3}). 
First, relative to the deterministic steady-state for
$\Delta F_0=20.5$, the dwell time of these solutions in the ice-free state is substantially longer.
In particular, note the significant difference 
in the date at which freeze up begins between $\Delta F_0=20.5$ and $21.0$. This highlights the fact that the exposed ocean is an effective heat reservoir and thus acts to prevent the formation of sea ice during the following winter season.  The effectiveness of this process depends on the time at which the ice disappears during the summer, and hence the time period that the open water is exposed to solar insolation \citep{MW:2012}.  Indeed, for all three values of $\Delta F_0$, sea ice only exists from early January to late May or early June, reflecting the time it takes to remove the stored heat from the mixed layer and bring it to the freezing temperature.  
Thus, the concave potentials in January represent the onset of heat loss from outgoing longwave radiative flux. 
As the ice becomes thinner, the curvature near the origin increases. By March, the longwave stabilization weakens and, 
particularly at $\Delta F_0=22.0$, the sea ice-albedo-feedback is already operative, which is reflected in the negative slope on the positive side of the potential  
(red curve). As the summer approaches, in all cases the ice-albedo-feedback strengthens and its magnitude increases 
with $\Delta F_0$, as seen through the changes in the slope on the positive side of the potentials. The two main competing physical processes, 
the longwave stabilization and the ice-albedo-feedback, are enhanced substantially during very short time-periods. Thus, 
the sensitivity of the system response to stochastic forcing increases.

\begin{figure}[H]
\centering
\includegraphics[angle=0,scale=0.25,trim= 0mm 0mm 0mm 0mm, clip]{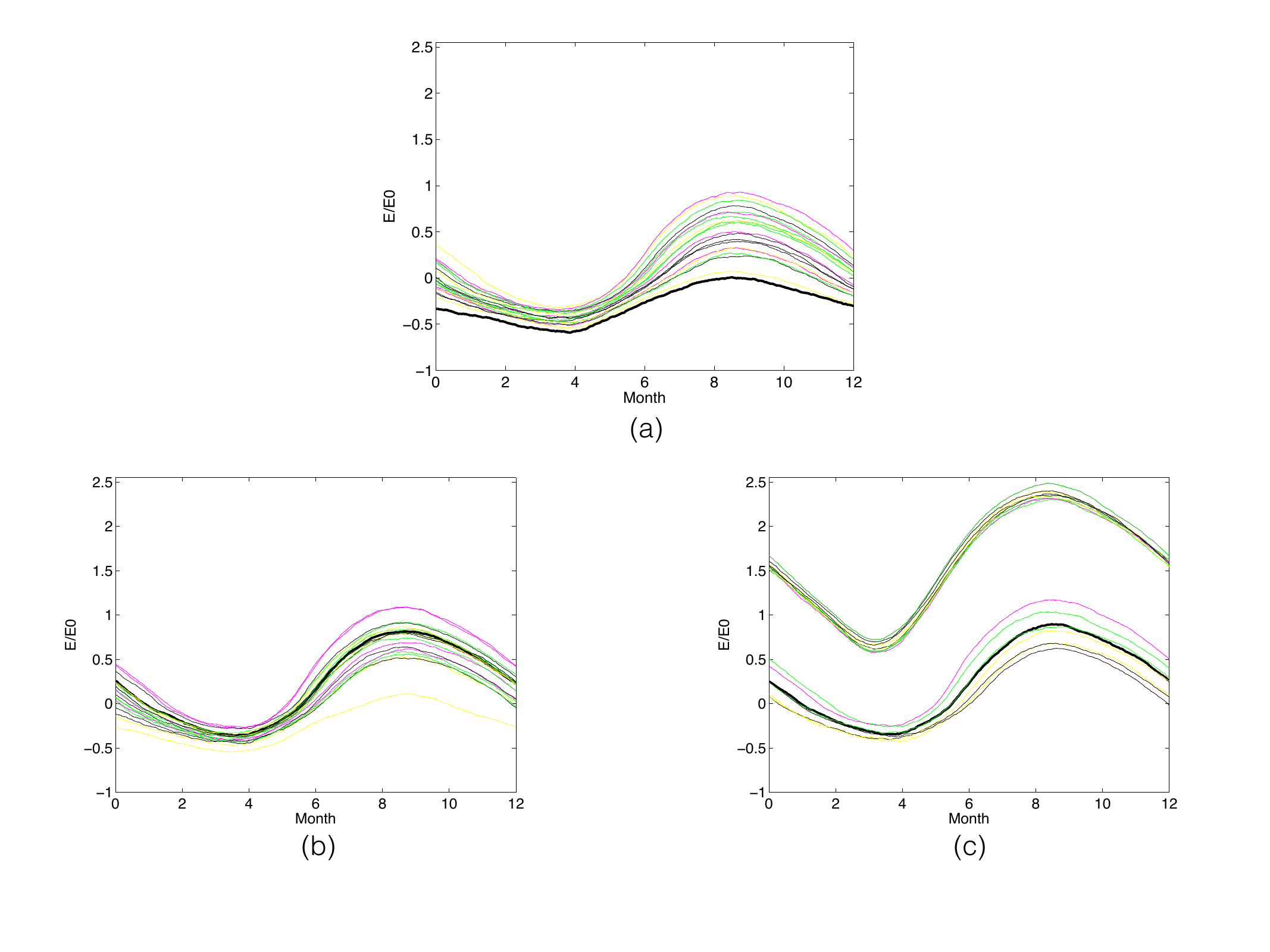}
\caption{Several realizations of the seasonal cycle of the stochastic solutions with three different values of $\Delta F_0$; (a) $21.0$, (b) $21.5$, (c) $22.0$.
         The thick black lines represent deterministic stable seasonal cycles of sea-ice-thickness. The other lines show 
         different realizations of the stochastic solutions.}
\label{fig:sample_path3}
\end{figure}

The striking behavior that emerges as $\Delta F_0$ approaches, but is still less than that for the deterministic saddle-node bifurcation, is seen in the stochastic paths of the seasonal cycle in 
Fig.~\ref{fig:sample_path3}.  For example, some stochastic paths for $\Delta F_0 = 21.5$ shown in Fig.~\ref{fig:sample_path3}(b) exhibit seasonal cycles at the extremes that are both barely seasonal ice-states, with small periods of either winter ice or ice free summers, thereby reflecting the deterministic transition.  With only a slight increase in $\Delta F_0$ the hysteresis of the deterministic backbone emerges with a two-state stochastic system in which seasonal ice and ice free states {\em coexist}, as seen in Fig.~\ref{fig:sample_path3}(c).  Importantly, this behavior would manifest itself in a transition with long dwell times in one of these two states and abrupt transitions between them. 

\section{\label{sec:resultsmult} Results : Comparing Multiplicative and Additive Noise}

As was done for additive noise, here we analyze the statistical properties of perennial and seasonally varying ice states separately.  
A physical origin of multiplicative noise are the fluctuations in the surface pressure field, which can be treated as a Gaussian random variable \citep{ThorndikeJAS:1982, Agarwal2017}.  This variability influences for example the ice transport from Fram Strait. Clearly, however, there are many other possible sources of noise.  As in the case of additive noise, we also use the ability to compare our analytical solutions with the numerical results, in the deterministic regime of perennial ice states where our perturbation theory is valid, as a well defined test bed of the numerical approach.

\subsection{\label{sec:perennial} Perennial ice states}

Stochastic paths are examined for all four cases of additive (CA, SVA) and multiplicative (IM, SM) noise.  For an objective comparison among the four cases, we generate the stochastic paths using the same random number at each time step drawn from a normal distribution with zero mean and standard deviation $\sqrt{\Delta t}$.  Therefore, the difference between the cases is intrinsic rather than arising from the randomness of the noise-forcing.  Overall, the stochastic solutions are well approximated by a Gaussian variable with a seasonally evolving standard deviation.

There is no substantial difference between SVA, IM and SM, but these differ from CA, which has a smaller variability.  This is intuitive, because the larger the magnitude of the noise-forcing during winter, the more effective it is in generating variability for SVA, IM and SM than in the case of a seasonally constant noise magnitude.   According to our perturbation  theory \citep{MW:2013}, 
all three cases have the same solution to first-order and are Gaussian variables with a standard deviation 
determined by the combination of the stability of the deterministic seasonal cycle and the noise-amplitude. The difference between the multiplicative noise characteristics of IM and SM appears at second order in perturbation theory.  The gap between their trajectories represents the intrinsic difference between It\^{o} and Stratonovich calculus. For example, this can be seen as a shift of the stochastic mean due to the cumulative effect of the noise-forcing represented explicitly in Stratonovich calculus. In this model the effect is always negative, the origin of which is the deterministic drift term that distinguishes the two calculi \citep{MW:2014}, and hence multiplicative noise generates more sea ice.

\begin{figure}[H]
\centering
\includegraphics[angle=0,scale=0.26,trim= 0mm 0mm 0mm 0mm, clip]{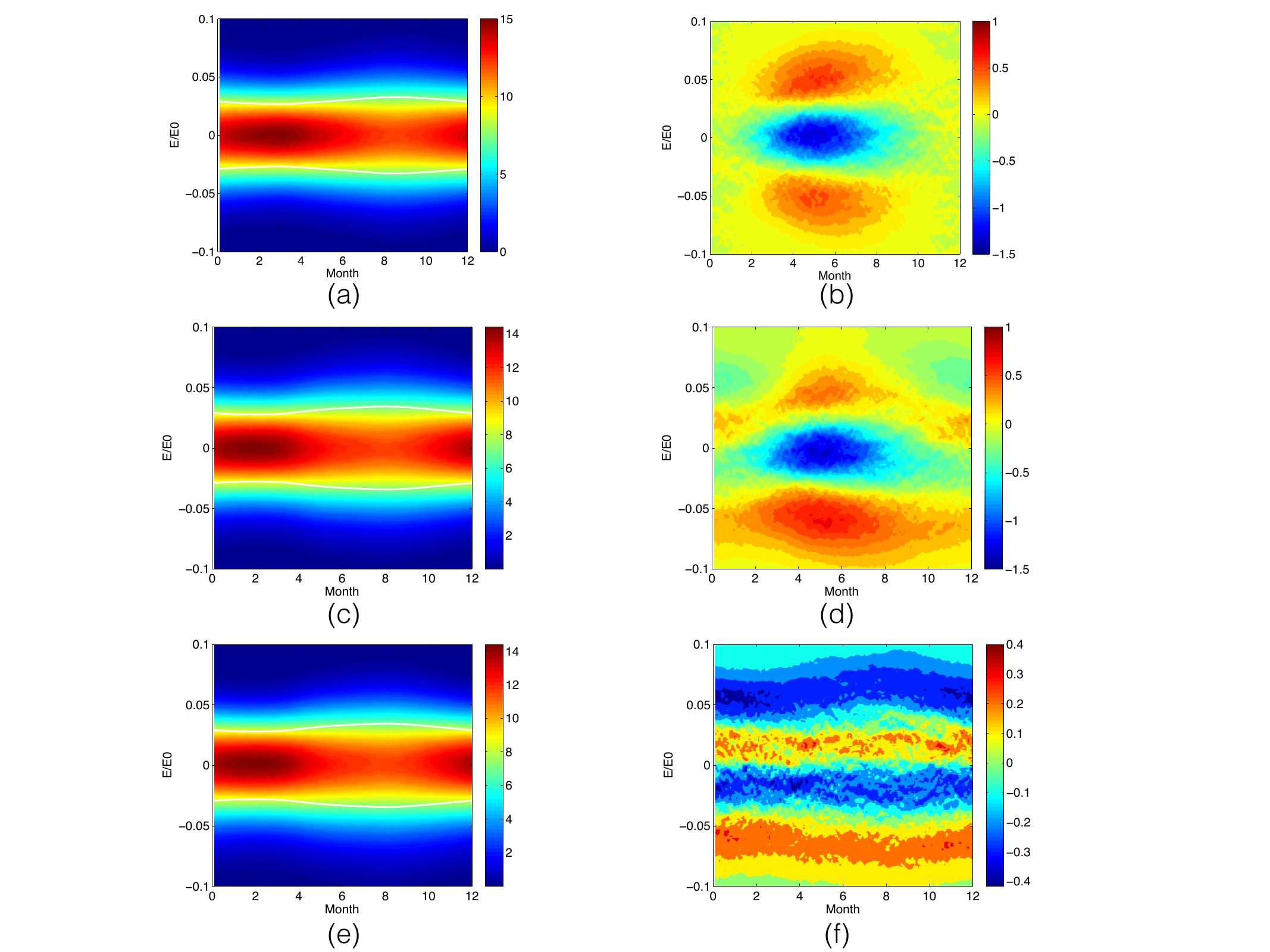}
\caption{Seasonal evolution of PDFs for CA (a), SVA (c) and SM (e) and the difference of the PDF, between SVA and CA (b), between SM and CA (d) 
         and between SM and SVA (f) are shown when $\Delta F_0 = 10.0$, in the perennial state of the deterministic system. 
         The X-axis is the the month of the year from January to December and the Y-axis the rescaled
         sea ice energy as in the previous figures.  The probability density is shown by the color scheme, where red represents higher values. The deterministic seasonal cycle is indicated by $E/E_0 = 0$, and the two white lines centered around $E/E_0 = 0$ indicate the standard deviation of the 
         stochastic solutions.}
\label{fig:pdf_seasonal01}
\end{figure} 

Clearly, because the solutions are periodic the PDFs change continuously during the year. To demonstrate this change we use the contour diagram shown in 
Fig.~\ref{fig:pdf_seasonal01}, with CA, SVA and SM in \ref{fig:pdf_seasonal01}(a), (c) and (e) respectively, for $\Delta F_0 = 10$. They are quite similar in the sense that the PDFs
are broad during summer and become narrow during winter, which is well explained by the two main competing effects of sea ice-albedo-feedback and longwave 
stabilization. The difference between pairs of these PDFs is shown in \ref{fig:pdf_seasonal01}(b),(d) and (f).
The difference between SVA and CA shown in (b) is characterized by the negative region near zero (blue) flanked by the positive regions, 
which shows that SVA has a wider PDF structure than CA. Note that this effect is particularly strong 
near the end of April, right before the sea ice-albedo-feedback starts to become active. We see that the noise-magnitude for SVA
is larger than CA during winter when the larger variability due to sea ice export is important, after which the sea 
ice albedo feedback becomes dominant. 
The comparison of SM and CA shown in (d) differs from that between SVA and CA in that the center of the negative region becomes more negative and the the positive region on the negative energy side  is more pronounced. This qualitative difference becomes more striking in (f), which shows that the center of the PDFs for SM become more negative
and more negatively skewed.

As $\Delta F_0$ increases from $10.0$, the competition between the sea ice-albedo-feedback and the longwave stabilization is amplified. 
Slightly thinner sea ice at the end of summer 
experiences increased longwave stabilization, which is effective throughout the following winter. At the same time, the magnitude of the noise-forcing decreases because
it is proportional to sea ice thickness, decreasing the overall variability. 
The imbalance between the longwave stabilization and the sea ice-albedo-feedback increases when $\Delta F_0 = 15.0$.  However,  a further increase in $\Delta F_0$ intensifies the sea ice albedo feedback thereby increasing the overall variability of the stochastic model. 

\begin{figure}[H]
\centering
\includegraphics[angle=0,scale=0.26,trim= 0mm 0mm 0mm 0mm, clip]{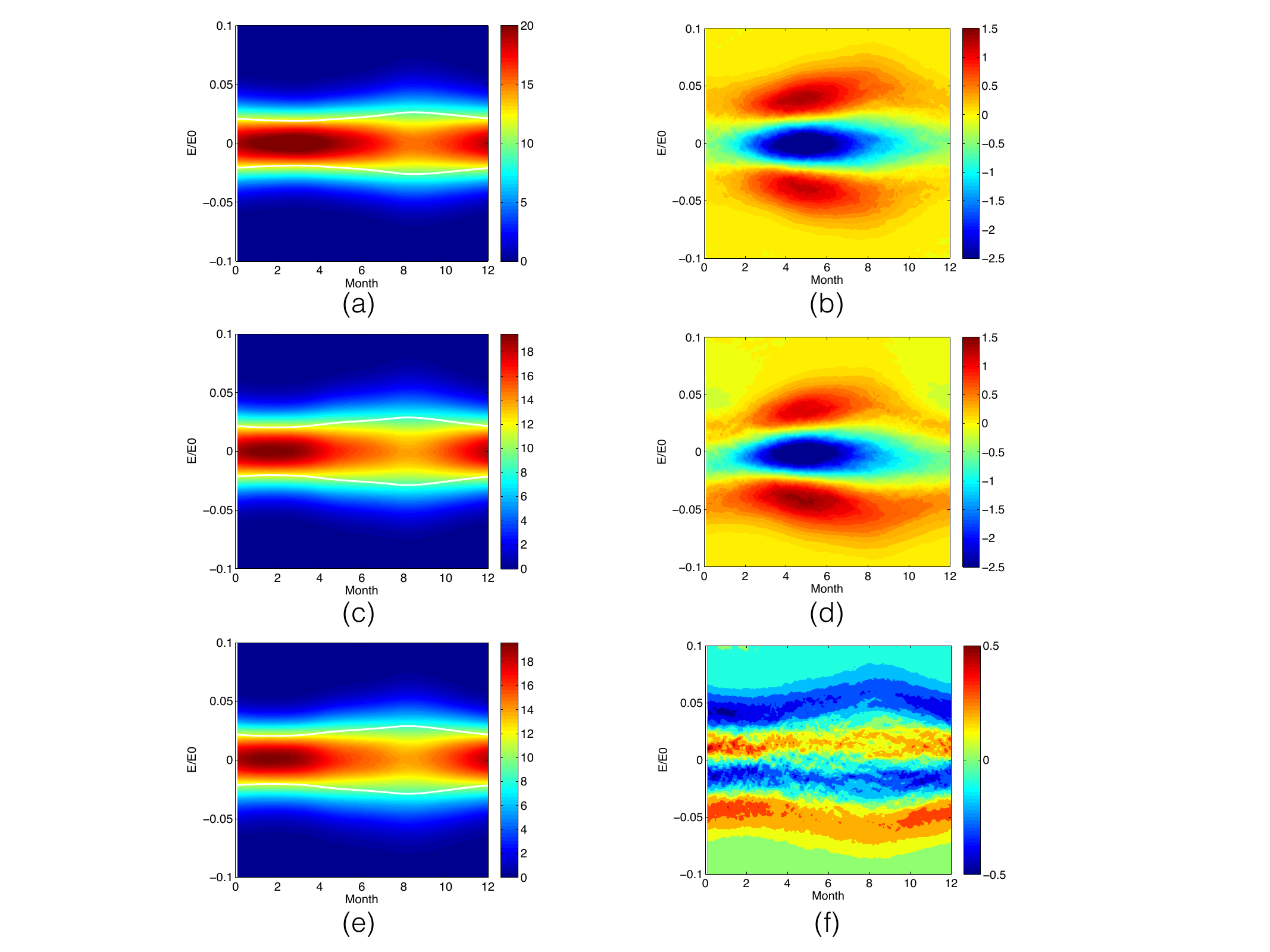}
\caption{Same as Fig.~\ref{fig:pdf_seasonal01} except that $\Delta F_0 = 15.0$.}
\label{fig:pdf_seasonal02}
\end{figure}

The contour diagram for $\Delta F_0 = 15.0$ is shown in Fig.~\ref{fig:pdf_seasonal02}. The individual PDFs for each case are nearly indistinguishable from each other, so 
we must examine the differences between them. We see from Figs.~\ref{fig:pdf_seasonal02}(b) and (d) that
the negative region around $E/E_0 = 0$ and the two positive regions flanking it represent the increasing breadth of the PDFs for SVA and SM  
relative to those for CA. 
The asymmetry associated with the multiplicative noise effect is shown in Figs.~\ref{fig:pdf_seasonal02}(d) and (f), with the increasingly 
darker red for $E/E_0 < 0$ and the overall negative shift of the PDFs for SM.

Having now examined $\Delta F_0=10$ and $15$, we can intuit that a further increase in $\Delta F_0$ will enhance the difference between CA and the other cases. 
We expect that the seasonal variation of the noise magnitude will generate larger variability
and this will couple to the increased influence of the sea ice-albedo-feedback during summer.
However, as the ice thins, so too will the impact of multiplicative noise, although the relative magnitude of the different contributions to the
overall variability are difficult to quantify. For example, as $\Delta F_0$ increases the stability of the ice cover weakens,
which provides the basis for the enhanced influence of stochastic forcing, but at the same time the 
magnitude of the noise-forcing decreases.

\begin{figure}[H]
\centering
\includegraphics[angle=0,scale=0.26,trim= 0mm 0mm 0mm 0mm, clip]{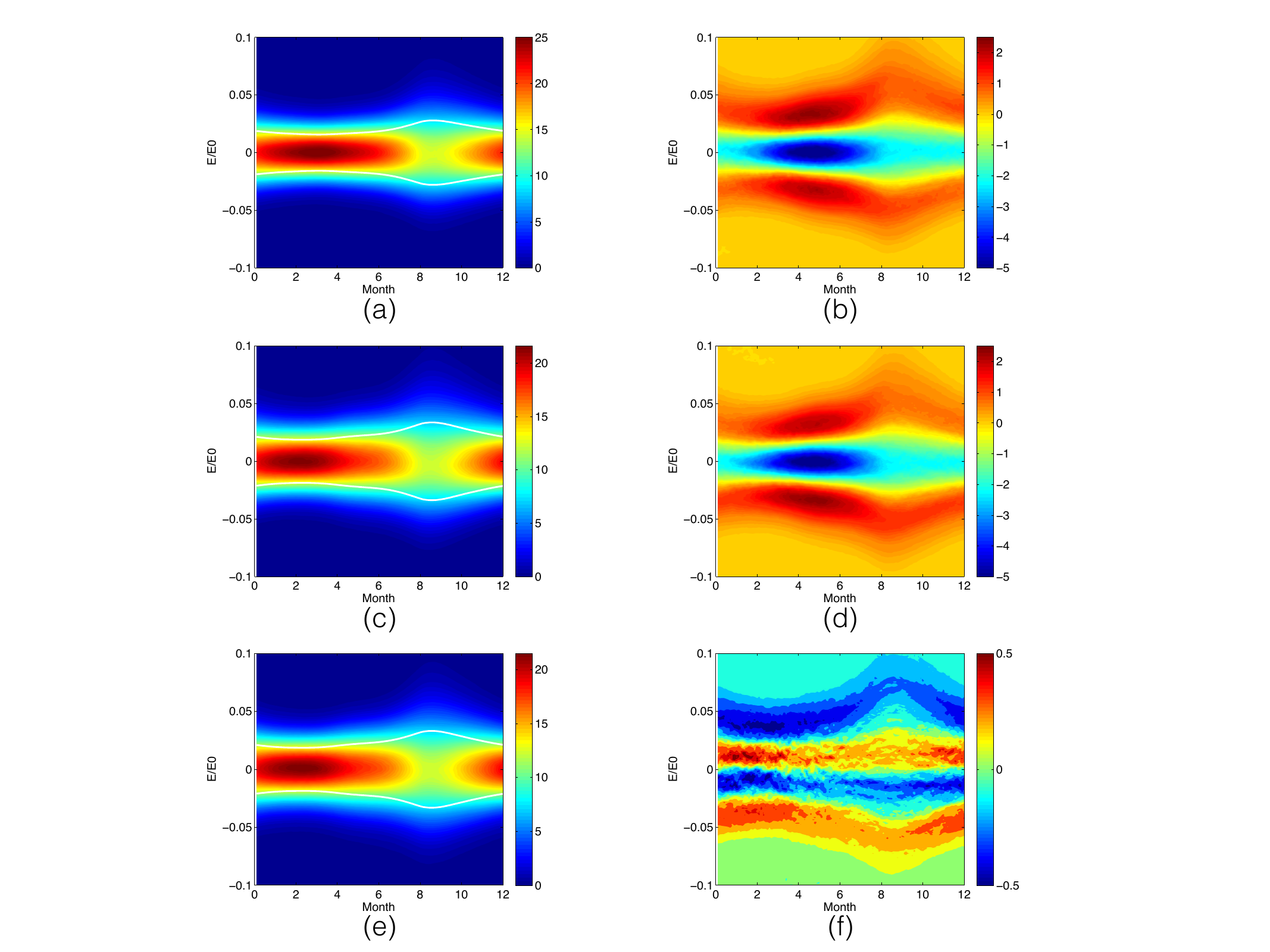}
\caption{Same as Fig.~ \ref{fig:pdf_seasonal01} except that $\Delta F_0 = 18.0$.}
\label{fig:pdf_seasonal03}
\end{figure}

For $\Delta F_0=18$ the PDFs of the stochastic solutions start to change dramatically, their spread around the deterministic seasonal cycle showing a strong seasonal dependence, as seen in the contour diagram of Fig.~\ref{fig:pdf_seasonal03}. 
Figs.~\ref{fig:pdf_seasonal03}(a), (c) and (e) show that the spread changes 
dramatically during the year, particularly at the end of a summer, where the standard deviation 
reaches a maximum.  Again, the sea ice-albedo-feedback is one of the principal contributors to the 
stochastic solution structure. The substantial  difference between CA and SVA and SM is shown in Figs.~\ref{fig:pdf_seasonal03}(b) and (d). 
The breadth of the PDFs due to the seasonal variation 
of the magnitude of the noise is exhibited again via the negative region centered around zero, flanked by the two positive regions. 
The temporal influence of the noise is such that its amplitude saturates in March, 
but the negative region appears later, between April and May.
The multiplicative noise effect shown in Fig.~\ref{fig:pdf_seasonal03}(f) is somewhat diminished relative to $\Delta F_0=10$. 
In particular, the positive regions (red) on the negative (lower) side
show that the negative tail of the PDFs is weaker than in the case with $\Delta F_0=10.0$.

In summary, as $\Delta F_0$ increases from $10.0$ to $19.0$, the deviation of the stochastic mean from the deterministic seasonal cycle changes from negative to positive for all the four cases, the difference between CA and SVA becoming larger with $\Delta F_0$.  The noise-forcing induced by the variability of sea ice export provides two important factors controlling the statistics of the stochastic solutions; 
(1) the seasonal change of the noise-magnitude and (2) the effect of multiplicative noise. 
The larger the magnitude of the noise near the end of winter the more effective it is in generating
increased variability of sea ice energy, and this becomes more important as $\Delta F_0$ increases. 
The effect of multiplicative noise, which always reduces the stochastic mean and the skewness, is stronger for lower values of $\Delta F_0$ 
because the noise-magnitude is proportional to the sea ice thickness. 
We end this section by noting that the approximate analytical solutions match well with the numerical solutions
suggesting that further research regarding the perennial ice states may be fruitfully explained using approximate methods \citep{MW:2013}.

\subsection{\label{sec:seasonal} Seasonally-varying states}

As $\Delta F_0$ increases, the deterministic dynamics predicts a reversible transition from perennial to seasonal ice, where sea ice vanishes during summer and grows back during winter.  It is notable that on the annual time scale the observed ice extent is a white noise signal \citep{Agarwal2012}, and we find here that the
seasonal states undergo dramatic fluctuations during the year.
Thin ice exposed to strong shortwave radiative flux during early summer melts quickly due to the sea ice-albedo-feedback. 
As winter approaches, 
thin ice forms from the open-ocean and then grows rapidly due to the strength of the longwave radiative heat loss. 
Regardless of the structure of the noise, its effect is to generate large variability around the deterministic seasonal cycle. Unfortunately, as mentioned above and previously \citep{MW:2013}, analytical solutions are not yet in hand for this regime. Nonetheless, the logic found in studying
 the perennial ice state acts as a framework for understanding stochastic solutions in the seasonal case. 

\begin{figure}[H]
\centering
\includegraphics[angle=0,scale=0.25,trim= 0mm 0mm 0mm 0mm, clip]{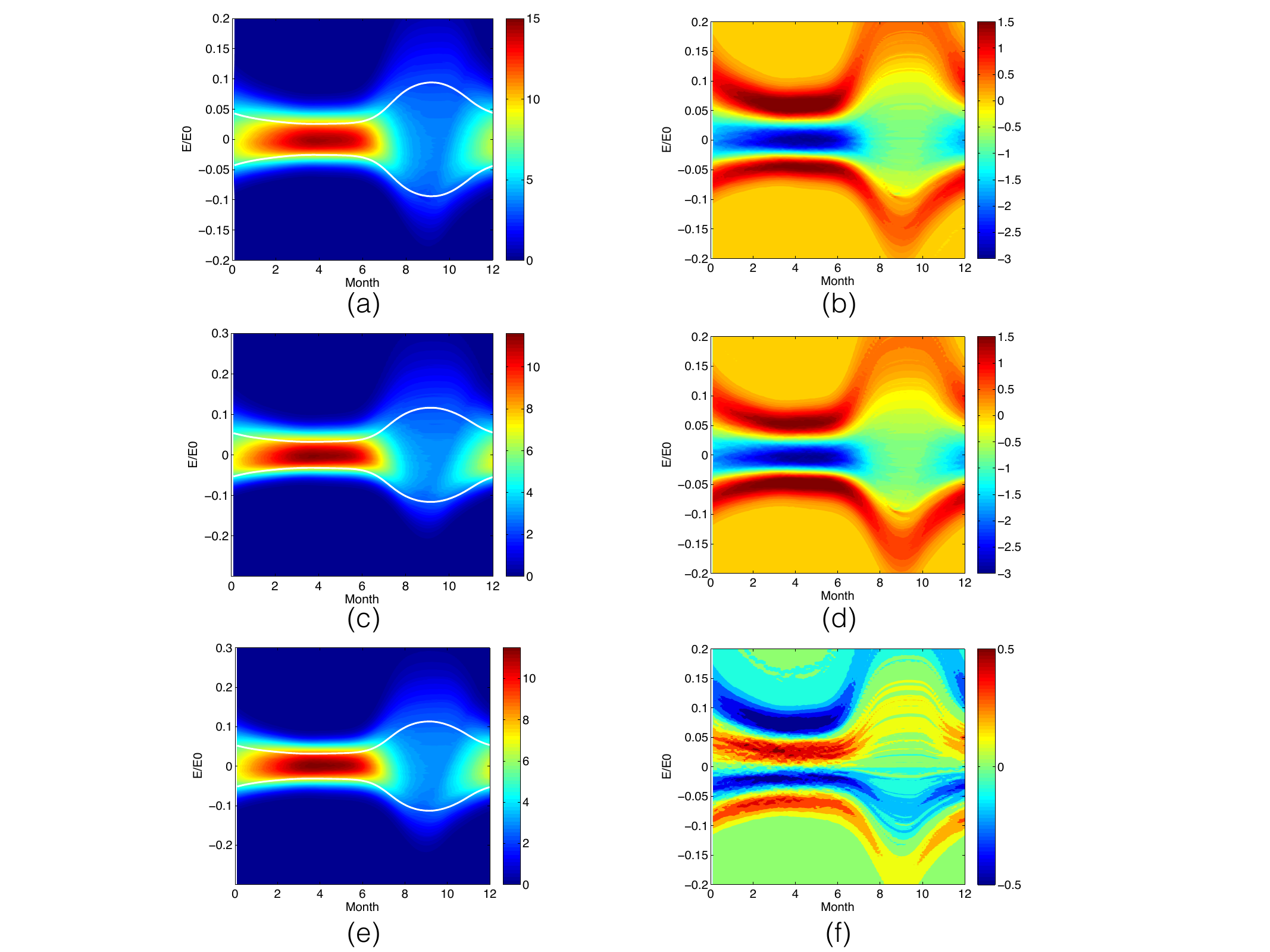}
\caption{Same as Fig.~ \ref{fig:pdf_seasonal01} except that $\Delta F_0 = 20.0$.}
\label{fig:pdf_seasonal04}
\end{figure} 

The continuous evolution of the PDFs over the year for all of the noise cases when $\Delta F_0=20.0$ is shown in Fig.~\ref{fig:pdf_seasonal04}. 
Even though the overall magnitude of the noise is smaller than 
that for the perennial sea ice states, the stochastic variability is even larger, which reflects the reduced stability of the system.
In the constant additive noise case,
we found that the PDFs at the same $\Delta F_0$
have positive tails due to the increased seasonal influence of the sea ice-albedo-feedback.

A key common characteristic of the PDFs is the distinct difference between summer and winter. 
The standard deviations (the two white lines) exhibit a dramatic change from winter to summer. Accordingly,
the shape of the PDFs also changes from sharply peaked to broad and the positive tails extend further in the positive sense during summer. 
These general characteristics are seen in CA, SVA and SM. The difference between SVA and CA, shown in Fig.~\ref{fig:pdf_seasonal04}(b), or between SM and CA, shown in Fig.~\ref{fig:pdf_seasonal04}(d), 
is qualitatively similar to the warmer (larger $\Delta F_0$) perennial ice states that exhibited continuous broadening. 
The difference between SVA and SM, shown in Fig.~\ref{fig:pdf_seasonal04}(f), is also similar to these previous cases, exhibiting 
a negative shift of the PDFs due to the drift term.

As $\Delta F_0$ increases slightly above $20.0$ we find large difference between the solutions. First,
the deterministic seasonal cycle changes rapidly with an increase in $\Delta F_0$ in this regime, for example, the open-ocean state persists much longer. Moreover, the sea ice-albedo-feedback becomes more sensitive to negative energy perturbations, which means that stochastic forcing generates more sea ice. 
Because the ice is thinner, the noise-amplitude is smaller and the variability for all four cases decreases.  

\begin{figure}[H]
\centering
\includegraphics[angle=0,scale=0.25,trim= 0mm 0mm 0mm 0mm, clip]{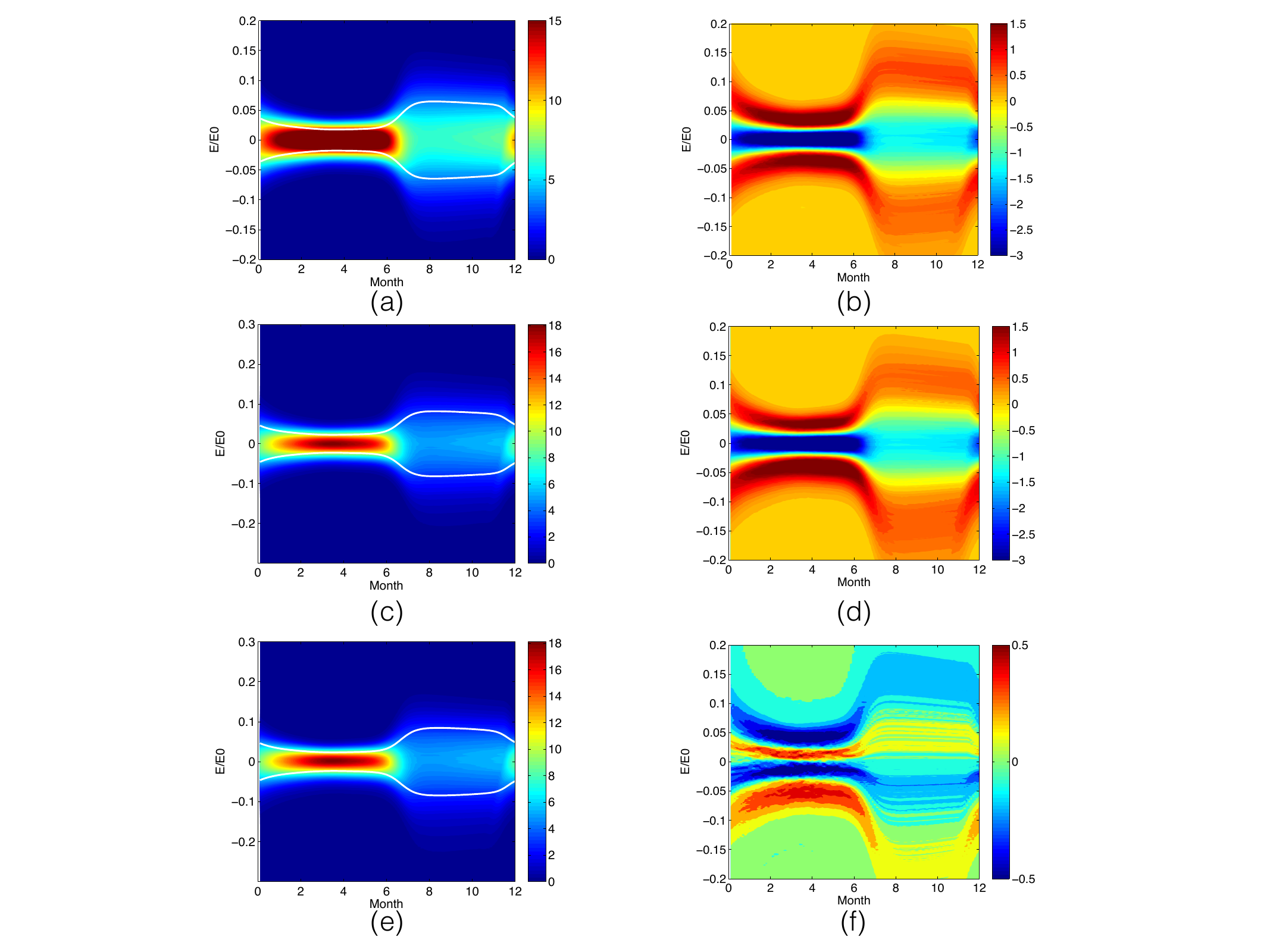}
\caption{Same as Fig.~ \ref{fig:pdf_seasonal01} except that $\Delta F_0 = 20.5$.}
\label{fig:pdf_seasonal05}
\end{figure} 

All of the PDFs for $\Delta F_0=20.5$ have negative tails (Fig.~\ref{fig:pdf_seasonal05}),
which is explained by the increased sensitivity of the albedo feedback to negative energy perturbations; 
the growth rate for a negative perturbation during summer is larger than that for a 
positive one. The qualitative consistency with the constant noise case is due to the decreased magnitude of the overall noise-forcing during the year. 
The seasonal variation of the noise-forcing and the effect of multiplicative noise do not make a significant difference. Thus, as expected, the PDFs during summer have broader negative tails. 

One of the most important characteristics of the CA, SVA and SM cases is the clear contrast between summer and winter, as seen in Figs.~\ref{fig:pdf_seasonal05}(a),(c) and (e), the origin of which is the dramatic change in the seasonal stability of the ice cover. The difference between CA and SVA shown in Fig.~\ref{fig:pdf_seasonal05}(b) is similar to the previous cases, 
with seasonal broadening and contraction. The PDFs for SM become more negative than those for SVA, shown as an increase in the red intensity straddling $E/E_0 = 0$ in Figs.~\ref{fig:pdf_seasonal05}(d) and (f).

The statistical moments are sensitive to small changes in $\Delta F_0$ in the seasonal state.
First, the standard deviation is slightly smaller than that at the lower $\Delta F_0$, which is due to the decreased noise-amplitude associated with the overall decay of the ice cover. Contrary to the sharp decrease after the maximum, the standard deviation decreases slowly after reaching the maximum and then shows a sharp decline in approximately November. 
Recall that at this time there is open-ocean, which has a large sensible heat and must be cooled before freezing can begin. 
After the ice forms, the strong longwave stabilization plays an important role in suppressing fluctuations. The deviation of the stochastic mean from the deterministic seasonal cycle is largely negative for all of the four cases. After the local maximum in June, a significant negative shift appears, which represents the sea ice-albedo-feedback 
being more sensitive to negative perturbations. Finally, near the transition
from the open-ocean to thin sea ice, there exists another local maximum. The first peak is associated with the sea ice-albedo-feedback in early summer and 
the second peak is the emergence of thin sea ice from open-ocean. When thin sea ice is generated, a perturbation
can be negative or positive. A positive perturbation leads to temporary melting of thin ice. 
The open-ocean has high heat capacity and thus stores substantial sensible heat, which always acts to delay the formation of thin sea ice.
Therefore, the system has positive asymmetry during the early stages of thin ice generation. 
After the ice is sufficiently thick, the strong longwave stabilization begins to control the stochastic solutions.

We summarize the statistics of the stochastic solutions in the seasonally varying state as follows. The standard deviation for SVA is larger than that for CA over the entire range of $\Delta F_0$, which was also seen for the perennial ice regime. The standard deviation is almost the same for SVA and SM, but a visible difference emerges near the deterministic saddle node bifurcation.
In the deviation of the stochastic mean from the deterministic seasonal cycle and the skewness, it is important to focus on the role of the sea ice-albedo-feedback near the deterministic transition from the perennial to the seasonally varying ice state. 
For $\Delta F_0 \approx 19.0$ (below the transition), both quantities are positive, which is associated with the nature of the ice albedo feedback, and is, as expected, enhanced for the SVA case. 
For the SM case, the negative multiplicative noise effect insures lower values than for the SVA case.  As discussed above, a slight increase in $\Delta F_0$ leads to a substantially different situation, as is evident in the negative deviation of the stochastic mean from the deterministic solution and the negative skewness. The increased sensitivity of the ice-albedo-feedback to a negative perturbation dominates the statistics immediately after the emergence of seasonally varying states. 
After passing through the transition, a sharp increase in the deviation of the stochastic mean from the deterministic solution and the skewness occurs until the deterministic saddle-node bifurcation to an ice-free state is approached.  Distinctions with the cases at lower $\Delta F_0$ include the skewness and the noise-magnitude for CA being larger than that for SVA, where the noise-magnitude is proportional to the 
ice thickness. Additionally, the skewness for SM is larger than that for SVA  and the skewness for the winter is larger than that for the summer.
The summer value is taken at the end of August when the open-ocean is stable relative to thin sea ice. At the end of March 
thin sea ice remains, which generates substantial sensitivity to perturbations.

\section{\label{sec:conclusion}Conclusion}

Using both analytical and numerical methods, we have studied the dependence of the solutions of a stochastic sea ice model on the external heat-flux $\Delta F_0$, which models
greenhouse gas forcing, for both additive and multiplicative noise.   Additive noise does not depend on the state of the system itself and is thus qualitatively and quantitatively distinct from multiplicative noise, which does depend on the state of the system.  Here, in the latter case we considered the variability of atmospheric forcing driving a variation of sea ice export as a key source of multiplicative noise, and hence the noise-forcing is linearly proportional to the sea ice thickness (or energy). 
The ensemble statistics of the system depend upon the stability and asymmetry of the underlying deterministic solutions and the magnitude of the noise-forcing. 
The stability and the asymmetry are principally determined by two main
processes; the ice-albedo-feedback and the longwave stabilization, which act asynchronously. 

We divided the analysis into the three regimes of $\Delta F_0$ associated with the steady state solutions of the deterministic system; perennial, seasonal and ice-free states as found by \citet{EW09}.  The deterministic perennial and seasonal states are separated by a reversible transition and the seasonal and ice-free states are delineated by a saddle-node bifurcation.  By introducing the concept and an ``ice-potential'', which describes the thermodynamic restoring forces in the system in a manner akin to a time-dependent Ornstein-Uhlenbeck process, we provide a relatively simple framework for interpreting the solutions.  

Because the underlying deterministic model is non-autonomous, so too is the stochastic model. 
When the noise-magnitude is small, and $\Delta F_0$ is such that the deterministic solutions are in the perennial state, we can  
compare numerical simulations with perturbative solutions derived previously \citep{MW:2013}.  This allows us to distinguish between the core nonlinear effects of the deterministic backbone of the model from those associated with noise-forcing at each order in the perturbative framework. We find a ``memory-effect'' whereby the intrinsic nonlinearity, asymmetry and stability characteristics of the interaction between the deterministic backbone and the noise allow fluctuations in ice energy from the early spring to accumulate and manifest themselves in the late summer.

We constructed and examined four variants of this noise-structure for a detailed comparison with the perturbative solution in the deterministic regime of stable perennial ice states.  The most general form of multiplicative noise-forcing is $\sigma {\cal R}(-E(t)) \xi(t)$, where $|\sigma|\ll1$ is the magnitude of the noise, 
$E(t)$ the sea ice energy and $\xi(t)$ is white noise.  Two cases were considered here, depending on the nature of the stochastic calculus;  It\^{o}-calculus (IM), which preserves the Martingale property, and Stratonovich-calculus (SM),  where the M denotes multiplicative. Because analysis of the properties of data alone is 
insufficient to determine which of the stochastic calculi is most appropriate for the task at hand, in the case of multiplicative noise we compare simulations from
both It\^o and Stratonovich calculi. The core reason for this is insufficient information regarding the difference in time-scale between noise forcing, inertia and/or feedbacks in the system, as is discussed in detail in \S \ref {sec:model_numeric}.\ref{sec:num} above.
The seasonally varying noise case (SVA), with noise amplitude $\sigma{\cal R}(-E_S)$ where $E_S(t)$ is the deterministic steady-state solution, examines the role of the seasonal change of the noise-amplitude. The constant additive noise case (CA) uses the seasonal average of $E_S(t)$, and thus has noise amplitude $\sigma\overline{\mathcal{R}(-E_S(t))}$, where the overbar denotes the seasonal time average.

In the perennial ice regime the difference between CA and SVA reveals the role of the seasonal variation of the noise-amplitude, and 
is detectable at first-order in perturbation theory, 
where the approximate solution is a Gaussian variable. As expected from the perturbation theory, the SM and IM cases exhibit no distinct 
difference with SVA at first-order. 
Rather, their differences are found at second order where non-Gaussian characteristics were predicted theoretically. 
Specifically, the difference between SVA and IM is seen in the skewness, due to the role of the effect of the multiplicative noise. 
The difference between IM and SM is due to the shift of the mean associated with the drift term in Stratonovich calculus. 

Even though the magnitude of the noise for the SVA case is larger (smaller) than that for CA during winter (summer), the seasonal standard deviation is larger. The overall behavior represents the confluence of the seasonal memory-effect with the variation of the noise-magnitude. As the external heat flux $\Delta F_0$
increases in this regime, the standard deviation decreases due to the decline of the noise-magnitude with the decline in ice-thickness, 
but increases again as $\Delta F_0$ increases further, due to the weakened stability associated with the ice-albedo-feedback. 
This change in the standard deviation with increasing $\Delta F_0$ explains the first-order solutions for all of the cases. 

For small $\Delta F_0$ the deviation of the stochastic mean from the deterministic seasonal cycle is negative  
for all of the cases, with SM having a larger deviation due to the nature of the mean shift induced by multiplicative 
noise. As $\Delta F_0$ increases further, 
the stochastic mean becomes larger than the deterministic seasonal cycle due to the ice-albedo-feedback. The 
difference between SVA and CA is particularly distinct, 
showing that the larger magnitude of the noise at the end of winter continues to impact the fluctuations of the 
sea ice energy during summer, which is the memory effect. 
The skewness behaves similarly to the deviation of the stochastic mean from the deterministic solution. 
The negative skewness for smaller values of $\Delta F_0$ increases sharply and becomes 
positive as $\Delta F_0$ increases. 
The effect of multiplicative noise in the SM and IM cases drives the sea ice energy towards negative values such that the skewness for 
these cases is smaller than that for SVA. 
The numerical results match the perturbation solutions nearly exactly, confirming the 
validity of the theoretical analysis in the perennial ice regime.

The seasonally varying states are clearly qualitatively and quantitatively different than the perennial states.  For example, 
the difference between SVA and CA is larger than in the perennial ice regime. Thus, quantitative estimation of sea ice variability in the seasonal state depends sensitively upon the detailed nature of the seasonality of the noise-magnitude. 
The controlling factor in the variability is the increased sensitivity of the ice-albedo feedback to negative energy (positive thickness) fluctuations near the transition from the perennial to the seasonally varying regime.  This {\em signed sensitivity} leads to both the deviation of the stochastic mean from the deterministic solution and the skewness having local 
minima near $\Delta F_0 = 20.5$, which is most pronounced in the SVA case.  These statistics pass through a smaller minimum in the SM and IM cases due to the nature of the multiplicative noise.  Finally, all of the statistical moments increase sharply as $\Delta F_0$ approaches the deterministic saddle-node bifurcation.

The central complexities of the evolution of the stochastic solutions as $\Delta F_0$ increases through the perennial and seasonally varying regimes of Arctic sea ice are best embodied in the evolution of the PDFs shown in Figs.~\ref{fig:pdf_seasonal01} , \ref{fig:pdf_seasonal02}, \ref{fig:pdf_seasonal03}, and \ref{fig:pdf_seasonal04}. Regardless of the regime, as $\Delta F_0$ increases the seasonality of the variability increases, but is maximal in the seasonal state.  The structure of other moments reveals the distinctions between additive and multiplicative noise, which becomes acutely important as the stability of the deterministic seasonal cycle weakens.  The asymmetry associated with the ice-albedo feedback response manifests itself in qualitatively unique ways when fluctuations are (not) tied to the ice energy/thickness in multiplicative (additive) noise.  There are a number of processes in which multiplicative noise is tied to observational reality, but as a general feature (independent of its origin) in this sort of a model it possesses some compelling features.  First, it captures leading order growth/decay of fluctuations, which we expect from general considerations of simple Langevin equations.  Second, in the case we considered here, as the ice cover is reduced then the fluctuations are less effective in impacting the state of the system, and this is clearly seen in the variability of the seasonal cycle and the nature of the memory effect.  It is thus of interest to systematically and explicitly incorporate stochastic effects in more complex models of sea ice, as is done in atmospheric models \citep{Dawson:2015}.  To this end, the framework provided here may be of use.

\acknowledgments
W.M. acknowledges a NASA Graduate Research Fellowship and a Herchel-Smith postdoctoral fellowship.  JSW acknowledges Swedish Research Council grant 638-2013-9243, a Royal Society Wolfson Research Merit Award and NASA Grant NNH13ZDA001N-CRYO for support.  Much of this work was completed at the 2015 Geophysical Fluid Dynamics Summer Study Program ``Stochastic Processes in Atmospheric \& Oceanic Dynamics'' at the Woods Hole Oceanographic Institution, which is supported by the National Science Foundation and the Office of Naval Research.  We thank many of the staff for comments and criticisms.

\end{document}